\documentclass[10pt]{article}
\usepackage[a4paper,bindingoffset=0.2in,%
            left=1in,right=1in,top=1in,bottom=1in,%
            footskip=.25in]{geometry}
\usepackage{graphicx,xcolor}
\usepackage{amsmath}
\usepackage{amsthm}
\usepackage{amssymb, latexsym, mathrsfs}
\usepackage{resizegather}
\usepackage{dsfont}
\usepackage{enumerate}
\usepackage{subcaption}
\usepackage{algorithm}
\usepackage{indentfirst}
\usepackage{color}
\usepackage{transparent}
\usepackage{url}
\usepackage{cite}

\usepackage[affil-it]{authblk}
\usepackage[colorlinks=true,citecolor=blue]{hyperref}
\begin{document}

     \title{\textbf{Decentralised $\mathcal{L}_1$ Adaptive Primary Controllers \& Distributed Consensus-Based Secondary Control for DC Microgrids with Constant-Power Loads}}
\author[1]{Daniel O'Keeffe\thanks{Research is supported by the Irish Research Council enterprise partnership scheme (Award No. EPSPG/2015/106) in collaboration with University College Cork, Ireland and United Technologies Research Centre Ireland Ltd.}\thanks{Email: \tt\small{danielokeeffe@umail.ucc.ie}; Corresponding author}}
\author[2]{Stefano Riverso\thanks{Email: \tt\small RiversS@utrc.ucc.com}}
\author[2]{Laura Albiol-Tendillo\thanks{Email: \tt\small AlbiolL@utrc.ucc.com}}
\author[1,3]{Gordon Lightbody\thanks{Email: \tt\small g.lightbody@ucc.ie}}
\affil[1]{\textnormal{Control \& Intelligent Systems Group, School of Engineering,
University College of Cork, Ireland}}
\affil[2]{\textnormal{United Technologies Research Centre Ireland Ltd, 4th Floor Penrose Business Centre, Cork, Ireland}}
\affil[3]{\textnormal{MaREI-SFI Research Centre, University College Cork, Ireland}}

     \date{\textbf{Technical Report}\\ March, 2018}

     \maketitle

     \begin{abstract}
Constant-power loads are notoriously known to destabilise power systems, such as DC microgrids, due to their negative incremental impedance. This paper equips distributed generation units with decentralised $\mathcal{L}_1$ adaptive controllers at the primary level of the microgrid control hierarchy. Necessary and sufficient conditions are provided to local controllers for overall microgrid stability when constant-power loads are connected. The advantages of the architecture over conventional heuristic approaches are: (i) scalable design, (ii) plug-and-play functionality, (iii) well defined performance and robustness guarantees in a heterogeneous and uncertain system, and (iv) avoids the need for online measurements to obtain non-\textit{a priori} system impedance information. The proposed primary control architecture is evaluated with distributed consensus-based secondary level controls using a bus-connected DC microgrid, which consists of DC-DC buck and boost converters, linear and non-linear loads. Stability of the overall hierarchical control system is proven using a unit-gain approximation of the primary level.

\end{abstract}

\newpage

\section{Introduction}

The increasing complexities associated with large-scale interconnected systems (LSiS) has led to the mass-adoption of decentralised control architectures over conventional centralised approaches \cite{Anuradha2013,Andreasson2016}. In fact, decentralised controllers have become the control standard for distributed autonomous power systems, also known as microgrids (mGs). The IEEE 2030.7 and 2030.8 standards outline guidelines for providing flexibility and plug-and-play (PnP) features in LSiS, which have recently become more imperative \cite{Stoustrup2009}. 

A key challenge for decentralised controllers in LSiS is to stipulate necessary and sufficient stability conditions. The stability of a mG can be compromised by the connection of constant-power loads (CPLs). In a small-signal sense, CPLs reduce system damping between individually stable systems by introducing negative-incremental impedance. Typically, mGs experience the effects of CPLs when tightly regulated motor drives are interfaced with inadequately damped power converters, or when multiple mGs are clustered together \cite{Shafiee2014}. Conventional criteria for stabilising mGs with CPLs have been reviewed in \cite{Riccobono2014,Vasquez2016}. Such approaches employ classical frequency domain analysis which is predicated on determining the impedance ratio between the interconnected power converters, known as the minor-loop gain, and are limited to unidirectional power flows, unlike state-space approaches. This requires \textit{a priori} knowledge of the total load number and the effective impedance value. Moreover, stability conditions are only sufficient, i.e. an unstable criteria can lead to a stable or an unstable system. The passivity-based criterion was proposed in \cite{Riccobono2012,Riccobono2013a} to overcome these restrictions by considering the overall system and measuring the effective impedance. Though the criterion is simple and practical, the tuning guidelines for local controllers are not provided. Ultimately, these designs are heuristic and not scalable, i.e. $N$ controllers may require retuning. As complex LSiS proliferate and become more heterogeneous, features such as flexibility, robustness to uncertainty and PnP operations will become increasingly more vital. Consequently, such restrictive approaches may become prohibitive.

Decentralised PnP control architectures have been proposed at the primary and secondary control levels of distributed generation units (DGUs) to guarantee overall voltage and current stability when power converters are plugged-in/out, irrespective of system topology. \cite{Riverso2015,Riverso2017a,Tucci2016c,Tucci2017g,Sadabadi2017a,Han2017}. To facilitate system scalability and reconfiguration, a control design orientated approach is adopted by using a load-connected model which treats loads as an exogenous disturbance. Kron reduction methods \cite{D??rfler2013} are subsequently used to map interconnections to more general network topologies such as bus-connected \cite{Tucci2017}. Among these works, only \cite{Sadabadi2017a} has considered CPL stabilisation within the PnP framework. Here, each local model uses a low-frequency approximation of the CPL and neglects high-frequency content. However, local stability conditions are only provided for DC-DC buck converters, while tests do not consider CPLs at all. Moreover, accurate knowledge of load power-ratings are required, which thus restricts compliance with flexible systems of different owners and stakeholders \cite{Katiraei2017}.

 Due to the increasing uncertainty within mGs, adaptive controllers have been implemented at the primary and secondary control levels \cite{Nasirian2014a,Josep2014,Vu2017}. However, these strategies are based on primary droop controllers, which are the conventional heuristic method for load-power sharing \cite{Lu2014,Simpson2013,Zhao2015}. These control schemes depend on specific mG models and topologies, and do not provide transient or steady-state robustness guarantees. Moreover, PnP operations are not considered. Recently, decentralised robust-adaptive control architectures have become attractive due to their guaranteed robustness during fast adaptation when compared to conventional model reference adaptive control (MRAC) architectures \cite{L12010,Gibson2012}.
 
 In \cite{OKeeffe2018e}, we developed a scalable decentralised $\mathcal{L}_1$ adaptive control ($\mathcal{L}_1$AC) architecture to augment the primary voltage controllers of DGUs. Steady-state robustness guarantees are provided in the presence of arbitrary topology, PnP operations, uncertain couplings and unknown load changes. However, global asymptotic stability (GAS) is achieved in a conservative fashion, while its computation is non-scalable. Subsequently, we formulated a distributed $\mathcal{L}_1$AC architecture to guarantee GAS in a scalable, PnP manner \cite{OKeeffe2018c}. Ultimately, these approaches are predicated on the ability to access hardware or software in order to augment or retrofit existing baseline controllers. In some applications this might not be available. 
 
 This paper extends our previous work by designing a purely adaptive architecture, and incorporating CPL stabilisation conditions for local controller tuning. Unlike the techniques of \cite{Riccobono2014,Vasquez2016,Lu2014a,Shafiee2014}, which implement stabilisers at the load-side converter, this paper follows \cite{Arcidiacono2012,Sadabadi2017a} whereby the DGU is equipped with each stabiliser. The rationale for this is that increasing damping by reducing the load-side control bandwidth impairs load-power quality. Moreover, it is the generators of the traditional utility grid which act to stabilise the overall system during load changes.
 
 The proposed architecture is implemented in a bus-connected DC mG, consisting of DC-DC boost and buck converters together with closed-loop speed controlled DC motors. Though the low-frequency CPL approximation of \cite{Sadabadi2017a} is also used, the low-pass filtering feature inherent in $\mathcal{L}_1$AC architectures, means that the high-frequency content of the CPL can be neglected without adversely affecting transient stability. Furthermore, due to the adaptive nature of the architecture, mapping bus-connected coupling parameters into the load-connected design model via Kron reduction can be avoided as long as the parameters are contained within the uncertainty subset.
 
 Finally, the proposed primary controllers are fitted with distributed consensus-based secondary controllers to achieve voltage restoration and load-power sharing objectives. Consensus algorithms have become popular in distributed systems as they can help decentralised controllers achieve centralised-like performance. Consensus allows nodes in a sparse communications network to construct a vision of the global system with limited information in a fault-tolerant manner. From the perspective of the secondary controllers, the primary level can be modelled as a unit-gain approximation \cite{Tucci2016}. Lyapunov functions are again used to demonstrate asymptotic stability using both approximations. The system's PnP capabilities and resiliency to communication faults is tested.

A version of this paper has been submitted to the 12$^{\textrm{th}}$ UKACC International Conference on Control.

\section{Constant-power load model}

 A bus-connected mG is a typical topology in automotive, marine and aircraft applications \cite{Luis2017} is shown in Fig. 1.
   \begin{figure}[!htb]    
    \graphicspath{ {Images/} }
    \centering
    \includegraphics[width=8.75cm]{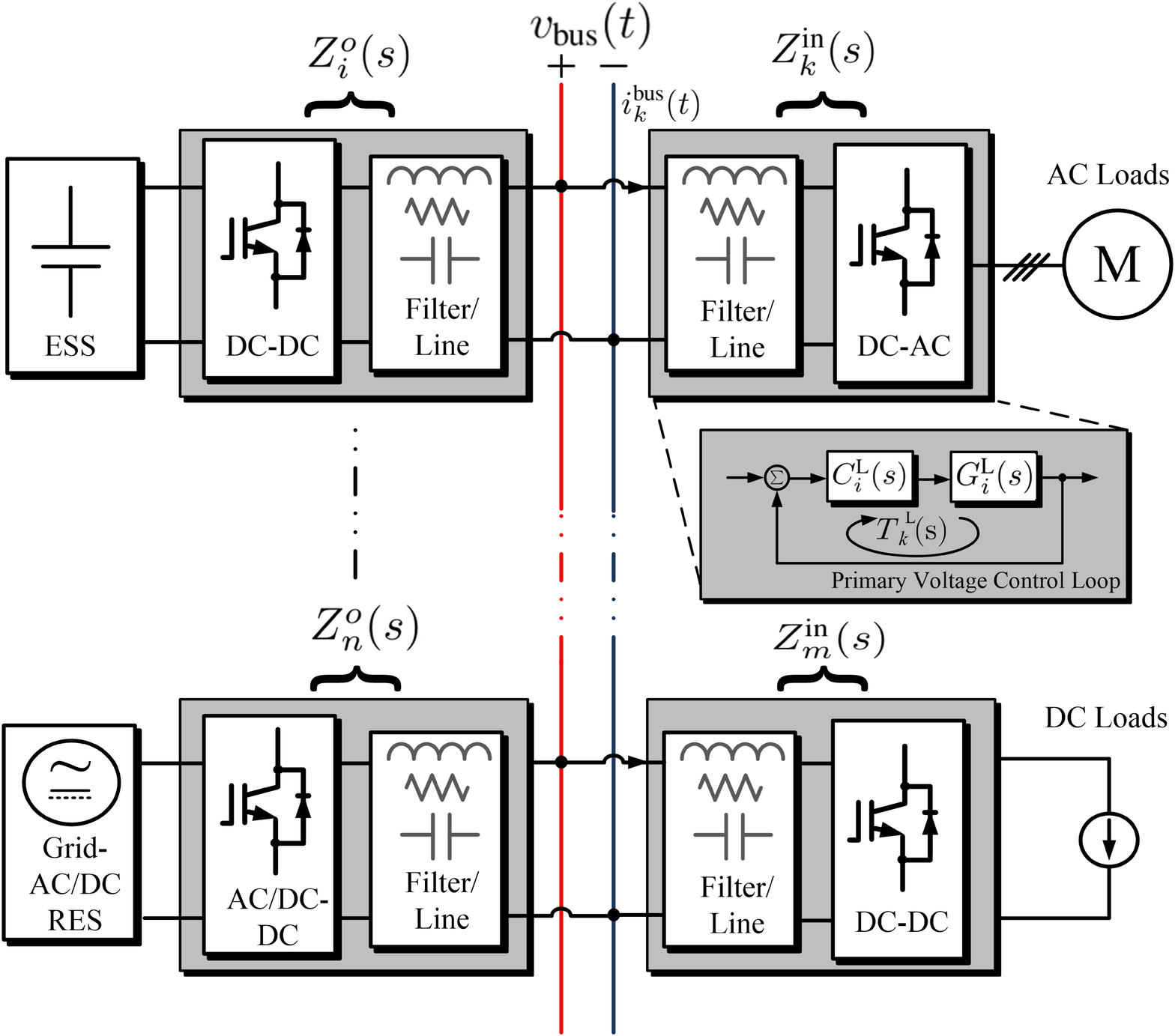}
    \caption{Typical DC microgrid configuration with $n$ generation-side and $m$ load-side converters.}
    \label{fig:ConvCct}
    \end{figure}
 
  Fig. 1 shows a mG bus formed via $n$ generation-side/DGUs and $m$ load-side converters. Each converter is locally fitted with voltage and current primary controllers. The primary control layer of the typical mG control hierarchy is required to provide fast and stable voltage and current performance in response to load changes and reconfiguration.
  
 This section derives the negative-incremental impedance model of load-side converters acting as CPLs, by decomposing each converter's closed-loop impedance into low and high frequency components using the Extra Element Theorem \cite{Riccobono2014}. At frequencies less than the closed-loop bandwidth i.e. where the controller loop-gain, $T_{k}^\textrm{L}(s)$ is large, and the controller works well, the closed-loop impedance of each load-side converter, $Z_{k}^{\textrm{in}}(s)$, approximates $Z_{k}^{\textrm{N}}(s)$, the forced impedance response. At frequencies greater than the closed-loop bandwidth, $Z_{k}^{\textrm{in}}(s)$ follows the open-loop impedance, $Z_{k}^{\textrm{D}}(s)$. This can be represented in terms of local admittance,
  \begin{equation}
  \frac{1}{Z_{k}^{\textrm{in}}(s)} = \frac{T_{k}^\textrm{L}(s)}{1+T_{k}^\textrm{L}(s)}\frac{1}{Z_{k}^\textrm{N}(s)}+\frac{1}{Z_{k}^\textrm{D}(s)}\frac{1}{1+T_{k}^\textrm{L}(s)}
  \label{eqn:CLZ1}
  \end{equation}
  where  $T_{k}^\textrm{L}(s) = C_{k}^\textrm{L}(s)G_{k}^\textrm{L}(s)$ assuming unity sensor and actuator gain, and $k \in \mathcal{M}=\{1,...,m\}$ denotes the set of load-converters, or number of CPLs. The overall admittance can be written as,
   \begin{equation}
 \mathbf{Y_{in}(s)} = \sum\limits_{i =1}^{m}\frac{1}{Z_{k}^{\textrm{in}}(s)}
   \label{eqn:CLZ2}
   \end{equation}
  A CPL can be expressed as,
 \begin{equation}
 P_k^\textrm{L} = v_\textrm{bus}(t)i_k^\textrm{bus}(t)
 \end{equation}
 where $P_k^\textrm{L}, v_\textrm{bus}$ and $i_k^\textrm{bus}$ represent the fixed power, and variable bus voltage and current. Within the closed-loop bandwidth, the voltage controller works to maintain a fixed load-power by adjusting its duty cycle to maintain a constant $v_k^\textrm{bus}(t)$, regardless of any input voltage disturbances from a mG bus. Typically, to exhibit good power quality, load-side converters are tuned to have very fast control-bandwidth. Therefore, if the bus voltage at the power converter's input terminals reduces, then $i_k^\textrm{bus}(t)$ will increase. This is the manifestation of negative-incremental impedance, which is analytically shown as,
 \begin{equation}
 \begin{aligned}
 Z_k^\textrm{N}(s) = R_k^\textrm{CPL} = \frac{\partial v_k^\textrm{bus}}{\partial i_k^\textrm{bus}}\rvert_{(V_\textrm{bus}, I_k^\textrm{bus})} = \frac{\partial}{\partial i_k^\textrm{bus}}\left(\frac{P_k^\textrm{L}}{I_k^\textrm{bus}}\right)\rvert_{(V_\textrm{bus}, I_k^\textrm{bus})}= -\frac{P_k^\textrm{L}}{{I_k^\textrm{bus}}^2}
 \label{eqn:negRinc}
  \end{aligned}
 \end{equation}
 For completion, the transfer function for the open-loop input impedance $Z_{k}^\textrm{D}(s)$ can be found in \cite{Vasquez2016}.
\section{Power converter models for DC microgrids}
This section defines the models for DC-DC source-side and load-side power converters. Boost converters, which increase input voltages, and buck converters, which decrease input voltages, are considered. Following the control design approach of \cite{Tucci2016c}, loads are treated as exogenous inputs in a load-connected topology. Fig.\ref{fig:ConvCct} shows each converter circuit topology with $RL$ power lines that connect converters to each other.
 \begin{figure}[!htb]    
  \graphicspath{ {Images/} }
  \centering
  \includegraphics[width=10.5cm]{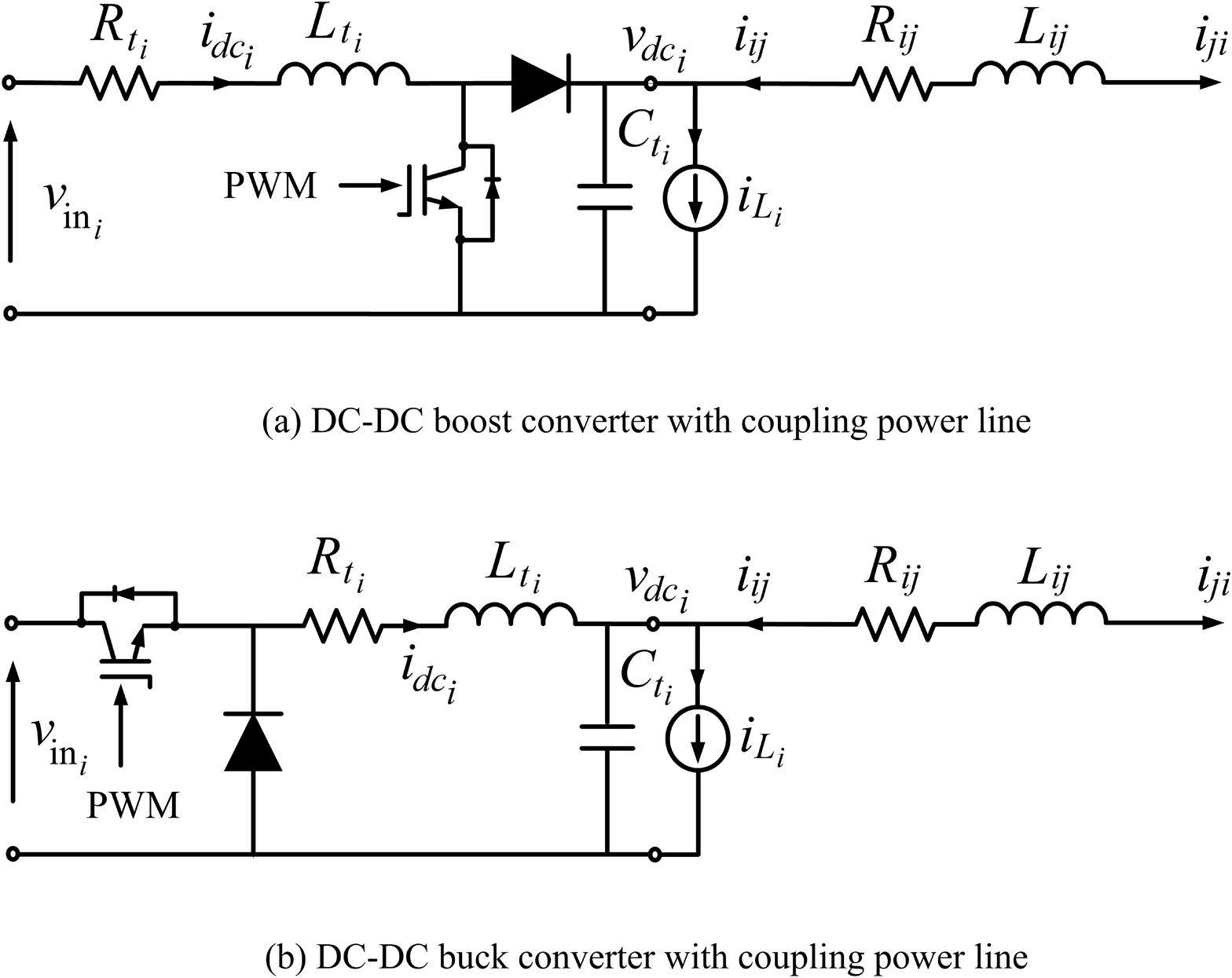}
  \caption{DC-DC power converter circuit topologies with power line connections to neighbours.}
  \label{fig:ConvCct}
  \end{figure}
  
Each converter can be modelled in small-signal state-space form as,
\begin{equation}
\Sigma_{[i]}^{\textrm{DGU}}:
\begin{cases}
\dot{x}_{[i]}(t) = A_{ii}x_{[i]}(t)+B_{i}u_{[i]}(t) + E_{i}w_{[i]}(t) + \zeta_{[i]}(t) \\
y_{[i]}(t) = C_ix_{[i]}(t)
\end{cases}
\label{eq:DGUSS}
\end{equation} 
where ${x}_{[i]}(t)= [\tilde i_{dc_{i}},\tilde v_{dc_{i}}]^T$, is the small-signal state vector, $u_{[i]}(t)$ is the small-signal control input, $w_{[i]}(t) = \tilde{i}_{L_{i}}$ is the small-signal load disturbance, represented as an exogenous input, $\zeta_{[i]}(t) = \sum_{j\in\mathcal{N}_i}A_{ij}x_{[j]}(t)$ represents coupling with $\Sigma_{[j]}^{\textrm{DGU}}$, and $(i,j)\in \mathcal{N}=\{1,...,n\}$ denotes the set of DGUs, and $j\in \mathcal{N}_i=\{1,...,n-1\}$ denotes the neighbour set of DGU $i$. 
From \cite{OKeeffe2018e}, the matrices of (\ref{eq:DGUSS}) for the boost converter are defined as,
\begin{equation*}	
\begin{aligned}
A_{ii} =
\left[ \begin{array}{cc}
-\frac{R_{t_{i}}}{L_{t_{i}}} & -\frac{(1-D_i)}{L_{t_{i}}}\\
\frac{(1-D_i)}{C_{t_{i}}} & -\sum_{j\in\mathcal{M}_i}\frac{1}{R_{ij}C_{t_{i}}}
\end{array} \right]
A_{ij}=\left[ \begin{array}{cc}
0 & 0 \\
0 & \frac{1}{R_{ij}C_{t_{i}}}
\end{array} \right]
B_{i} = \left[ \begin{array}{c}
\frac{V_{dc_{i}}}{L_{t_{i}}}\\
\frac{-I_{dc_{i}}}{C_{t_{i}}}
\end{array} \right]
E_i = 
\left[ \begin{array}{c}
0\\
\frac{-1}{C_{t_{i}}}
\end{array} \right]
C_i =
\left[ \begin{array}{cc}
0 & 1
\end{array} \right]
\end{aligned}
\end{equation*}
where $A_{ii} \in \mathbb{R}^{2\times2}$ is the state matrix; $A_{ij} \in \mathbb{R}^{2\times2}$ is the coupling matrix, $B_{i}\in \mathbb{R}^2$ is the input vector, and $C_i \in \mathbb{R}^{1\times2}$ is the output vector. Furthermore, $V_{dc_{i}} = \frac{V_{\textrm{in}_{i}}}{(1-D_i)}$, and $I_{dc_{i}} = \frac{V_{\textrm{in}_{i}}}{(1-D_i)^2R_{L_{i}}}$, where $R_{L_{i}}$ is the effective local load resistance at each converter's terminals. However, since the lad-connected topology treats loads as disturbance $R_{L_{i}}$ is unknown.

From \cite{Tucci2016c}, the matrices of (\ref{eq:DGUSS}) for the buck converter are defined as,
\begin{equation*}
\begin{aligned}
A_{ii}=
\left[ \begin{array}{cc}
-\frac{R_{t_{i}}}{L_{t_{i}}} & -\frac{1}{L_{t_{i}}}\\
\frac{1}{C_{t_{i}}} & -\sum_{j\in\mathcal{M}_i}\frac{1}{R_{ij}C_{t_{i}}}
\end{array} \right]
A_{ij}=\left[ \begin{array}{cc}
0 & 0 \\
0 & \frac{1}{R_{ij}C_{t_{i}}}
\end{array} \right]
B_{i} = \left[ \begin{array}{c}
\frac{1}{L_{t_{i}}}\\
0
\end{array} \right]
E_i = 
\left[ \begin{array}{c}
0\\
-\frac{1}{C_{t_{i}}}
\end{array} \right]
C_i =
\left[ \begin{array}{cc}
0 & 1
\end{array} \right]
\end{aligned}
\end{equation*}
Comparing both converter models, it is clear that the dynamics of the boost are dependent on its duty cycle operating point and is non-minimum phase (NMP) for output voltage control. The NMP action makes controller tuning more difficult, particularly when coupled to unknown power lines as addressed in \cite{OKeeffe2017a}. 

To introduce the CPL model of (\ref{eqn:negRinc}), the exogenous input is altered such that ${i}_{L_{i}} = {i}_{i}^{\textrm{CPL}}+{i}_{l_{i}}$,  where $\tilde{i}_{l_{i}}(t)$ represents the current disturbance due to non-CPL loads i.e. neighbouring DGUs or battery banks. The large-signal equivalent line resistances and currents at the load-connected terminals of each DGU must be mapped from the original bus-connected topology using Kirchoff's voltage and current laws, as in \cite{Riverso2017a};
\begin{equation}
\begin{aligned}
R_{ij} = R_iR_j\sum_{q=1}^{m}\frac{1}{R_q}, \forall j \neq i \\
{i}_{i}^{\textrm{CPL}} = \frac{\sum_{k=1}^{m}i_{k}^{\textrm{bus}}}{\left(R_k\sum_{q=1}^{m}\frac{1}{R_q}\right)} = \frac{\frac{1}{v_{\textrm{bus}}}\sum_{k=1}^{m}P_k^{\textrm{L}}}{\left(R_k\sum_{q=1}^{m}\frac{1}{R_q}\right)}
\label{eqn:bus2loadmap}
\end{aligned}
\end{equation}
where $R_i$ and $R_j $ are the line resistances that connect DGUs to the mG bus as in Fig. 1. Furthermore, as the mG bus voltage varies depending on the line resistances and the number of grid-forming DGUs present, $v_{\textrm{bus}}$ is represented in terms of the output voltage of each DGU. 
\begin{equation}
v_{\textrm{bus}} = \frac{\sum_{i=1}^{n}\frac{v_{dc_{i}}}{R_i}}{\sum_{k=1}^{m}\frac{1}{R_k^{\textrm{CPL}}}+\sum_{i=1}^{n}\frac{1}{R_i}}
\label{eqn:vbus1}
\end{equation}
Using (\ref{eqn:negRinc}) and substituting $\sum_{k=1}^{m}\frac{1}{R_k^{\textrm{CPL}}} =  -\frac{1}{v_{\textrm{bus}}^2}\sum_{k=1}^{m}P_k^{\textrm{L}}$ into (\ref{eqn:vbus1}) yields a quadratic solution for $v_{\textrm{bus}}$,
\begin{equation}
v_{\textrm{bus}} = \frac{\sum_{i=1}^{n}\frac{v_{dc_{i}}}{R_i} \pm \sqrt{\left(\sum_{i=1}^{n}\frac{v_{dc_{i}}}{R_i}\right)^2+4\sum_{k=1}^{m}P_k^{\textrm{L}}\sum_{i=1}^{n}\frac{1}{R_i}}}{2\sum_{i=1}^{n}\frac{1}{R_i}}
\end{equation}
Alternatively, if the upper-bound $R_k^{\textrm{CPL}}$ is known, i.e. within the bounds of adaptation, the effective CPL rating can be written as, 
\begin{equation}
P_{i}^{\textrm{CPL}} = \frac{\left(\frac{1}{R_k^{\textrm{CPL}}}+\sum_{i=1}^{n}\frac{1}{R_i}\right)\sum_{k=1}^{m}P_k^{\textrm{L}}}{\frac{1}{R_i}\left(1+\frac{1}{v_{dc_{i}}}\sum_{j\in\mathcal{N}_i}v_{dc_{j}}\right)\left(R_k\sum_{q=1}^{m}\frac{1}{R_q}\right)}
\end{equation}
Finally, ${i}_{L_{i}}$ must be linearised about the operating point as ${i}_{i}^{\textrm{CPL}} = \frac{P_{i}^{\textrm{CPL}}}{v_{dc_{i}}}$ is non-linear. This yields,
\begin{equation}
\tilde{i}_{i}^{\textrm{CPL}} = -\frac{P_{i}^{\textrm{CPL}}}{V_{dc_i}^2}\tilde v_{dc_{i}}
\end{equation}
As a result, the state matrix of the boost converter model becomes,
\begin{equation}
A_{ii}=
\left[ \begin{array}{cc}
-\frac{R_{t_{i}}}{L_{t_{i}}} & -\frac{(1-D_i)}{L_{t_{i}}}\\
\frac{(1-D_i)}{C_{t_{i}}} & -\frac{1}{C_{t_{i}}}\left(\sum_{j\in\mathcal{N}_i}\frac{1}{R_{ij}}-\frac{P_{i}^{\textrm{CPL}}}{V_{dc_{i}}^2}\right)
\end{array} \right]
\end{equation}
and the state matrix of the buck converter model becomes,
\begin{equation}
A_{ii}=
\left[ \begin{array}{cc}
-\frac{R_{t_{i}}}{L_{t_{i}}} & -\frac{1}{L_{t_{i}}}\\
\frac{1}{C_{t_{i}}} & -\frac{1}{C_{t_{i}}}\left(\sum_{j\in\mathcal{N}_i}\frac{1}{R_{ij}}-\frac{P_{i}^{\textrm{CPL}}}{V_{dc_{i}}^2}\right)
\end{array} \right]
\end{equation}
Ultimately, the low frequency approximation of the CPL can no longer be modelled as an exogenous input as it now directly influences the eigenvalues of both converters. This introduces greater uncertainty into the model as loads are unknown when considering flexible and heterogeneous DC mGs.  

\section{Decentralised $\mathcal{L}_1$ adaptive primary voltage control}
Over the last decade, efforts to improve the transient performance and robustness guarantees of conventional MRAC architectures with fast adaptation, led to the formulation of $\mathcal{L}_1$AC theory \cite{L12010}. This robust-adaptive control architecture achieves transient performance and bounded state and control signal guarantees by inserting a low-pass filter (LPF) at the input to both the plant and state-predictor of an indirect MRAC. 
The theory has been developed state and output feedback architectures for time-varying uncertainties and disturbances, unmodelled dynamics, time-delays, unknown input gains and non-linearities \cite{L12010}.

Centralised $\mathcal{L}_1$AC architectures have been notably successful in safety-critical applications including; sub-scale NASA aircraft auto-pilots \cite{Gregory2010}, manned aircraft \cite{Ackerman2017}, and unmanned water/aerial vehicles \cite{Svendsen2012,Michini2009}. In addition to our decentralised \cite{OKeeffe2018e} and distributed \cite{OKeeffe2018c} $\mathcal{L}_1$AC architectures, decentralised schemes were formulated in \cite{Yoo2010,DeL1AC}, and were implemented to augment aircraft baseline controllers \cite{Kumaresan2016}. Unlike \cite{Gregory2010,Ackerman2017,Svendsen2012,Michini2009,Yoo2010,DeL1AC,Kumaresan2016}, we consider unmatched interconnections and provide robustness to constant disturbances via integral action.

\subsection{Plant structure}
The plant has a known structure, but with unknown parameter values. The control objective is to design a bounded control input $u_{[i]}(t)$, such that $y_{[i]}(t) = C_ix_{[t]}(t)$ tracks a reference voltage with convergent state and bounded parametric errors in the presence of matched uncertainty and unmatched coupling and disturbances. Defining the new state,
\begin{equation}
\begin{aligned}
\dot{\xi}_{[i]}(t) = y_{[i]}^{\textrm{ref}}(t) - y_{[i]}(t) \\
\xi_{[i]}(t) = \int_{0}^{t}\left(y_{[i]}^{\textrm{ref}}(t) - y_{[i]}(t)\right) dt = \int_{0}^{t}\left(y_{[i]}^{\textrm{ref}}(t) - C_{i}x_{[i]}(t)\right) dt
\end{aligned}
\label{eqn:plant}
\end{equation}
where $y_{[i]}^{\textrm{ref}}(t)=V^{\textrm{ref}}_{dc_{i}}-V_{dc_{i}}$. \vspace{3mm} \newline 
\textbf{Remark 1.} \textit{Initially, the unmatched coupling term $\zeta_{[i]}$ is neglected to enable local decoupled design. Subsequently, the term is reintroduced when GAS conditions are provided. Moreover, as in \cite{OKeeffe2018e}, the integrator provides adequate robustness to unmatched constant disturbances, and therefore $\tilde{i}_{l_{[i]}}(t)$ is neglected.}   
 \vspace{3mm}
The augmented open-loop model, where a matched uncertainty term is introduced to represent parametric uncertainty in the dynamics of $\Sigma_{[i]}^{\textrm{DGU}}$ can be represented as,
\begin{equation}
 \bar{\Sigma}_{[i]}^{\textrm{DGU}}:
 \begin{cases}
  \dot{\bar{x}}_{[i]}(t) = \bar{A}_{ii}\bar{x}_{[i]}(t)+ \bar{B}_{i}\left(u_{[i]}(t) +\bar{\theta}_{[i]}(t)\hat{x}_{[i]}(t)\right) + Fy_{[i]}^{\textrm{ref}}(t) \\
 \bar{y}_{[i]}(t) = \bar{C}_i\bar{x}_{[i]}(t)
 \end{cases}
 \label{eq:L1SS2}
 \end{equation}
where $\bar{x}_{[i]}(t) \in \mathbb{R}^{3}$, is the system measurable state vector; $u_{[i]}^{\mathcal{L}_1}(t) \in \mathbb{R}$ is the control signal; $\bar{\theta}_{[i]}(t)\in \Theta \subset \mathbb{R}^3$ is the unknown parametric uncertainty vector, which belongs to the known uniformly bounded  convex set $\Theta$. The matrices of (\ref{eq:L1SS2}) are defined as,
\begin{equation*}
\begin{aligned}
\bar{A}_{ii}=
\left[ \begin{array}{cc}
A_{ii} & 0_{2\textrm{x}1}\\
-C_i & 0 
\end{array} \right]
\bar{B}_{i} = \left[ \begin{array}{c}
B_i\\
0
\end{array} \right]
F = \left[ \begin{array}{c}
0\\
0\\
1\end{array} \right]
\bar{C}_i =
\left[ \begin{array}{cc}
C_i & 0\end{array} \right]
\end{aligned}
\end{equation*}

\subsection{Control law}
The small-signal control input $u_{[i]}(t)$ for $\bar{\Sigma}_{[i]}^{\textrm{DGU}}$ is a fusion of a state-feedback controller and low-frequency banded uncertainty compensation signal, as defined in the Laplace domain, 
\begin{equation}
\mathcal{C}^{\mathcal{L}_1}_{[i]}: u_{[i]}^{\textrm{SF}}(t)+u_{[i]}^{\mathcal{L}_1}(t) = -\left(K_i\bar{x}_{[i]}(t) +C(p)[\hat{\theta}_{[i]}^T\bar{x}_{[i]}](t)\right)
\label{eqn:C_L1}
\end{equation}
where $K_i = [K_i^i, K_i^v, K_i^{\xi}]\in\mathbb{R}^{1\times3}$ is the state-feedback control gain vector, $C(p)$ represents a second 
-order Butterworth LPF\footnote{A second-order Butterworth LPF is chosen as it has a maximally flat pass-band, and the order must be at least equal to the order of the closed-loop plant to ensure a bounded control input - see section \ref{DesignConsiderations}}, and $p\triangleq \frac{d}{dt}$ represents the differential operator. The robustness of the $\mathcal{L}_1$AC is dependent on the LPF bandwidth $\omega_c$, as subsequently designed.
\vspace{-1mm}
\subsection{State-predictor}
The state-predictor generates an estimate of the system states. Thereafter, the adaptive law is used to asymptotically drive the uncertain plant dynamics to converge to the desired closed-loop dynamics of the state-predictor. Without loss of generality, the state-predictor formulation and the desired closed-loop dynamics are equal for all DGUs,
\begin{equation}
 \mathcal{E}_{[i]}:
 \begin{cases}
 \dot{\hat{x}}_{[i]}(t) = \hat{A}_{m}\hat{x}_{[i]}(t)+\hat{B}_m(u_{[i]}^{\mathcal{L}_1}(t)+\hat{\theta}_{[i]}^T(t)\hat{x}_{[i]}(t)) + Fy_{[i]}^{\textrm{ref}}(t)
\\
\hat{y}_{[i]}(t) = \hat{C}_{i}\hat{x}_{[i]}(t)
 \end{cases}
 \label{eq:L1SPSS2}
 \end{equation}
where $K_i$ renders $\hat{A}_{m} \triangleq  \bar{A}_{ii}^{\textrm{nom}} - \bar{B}_iK_i \in \mathbb{R}^{3\times3}$ Hurwitz, $\hat{A}_{m}$ is the design matrix that specifies the desired closed-loop dynamics, and $\bar{A}_{ii}^{\textrm{nom}}$ is the state matrix in which the designer estimates as the nominal dynamics without uncertainty in order to design the nominal control gains. 

\subsection{Adaptive law} 
The adaptive law generates an estimate of the plant uncertainties. Defining the state-error and parametric estimation error vectors as, $\tilde{{x}}_{[i]}(t) = \bar{x}_{[i]}(t) - \hat{x}_{[i]}(t)$ and $\tilde{{\theta}}_{[i]}(t) = \bar{\theta}_{[i]}(t) - \hat{\theta}_{[i]}(t)$, the state-error dynamics, used to drive the adaptive law, can be defined as,

\begin{equation}
\dot{\tilde{x}}_{[i]}(t) = \hat{A}_{m}\tilde{x}_{[i]}(t)+\hat{B}_{m}\tilde{\theta}_{[i]}(t)\hat{x}_{[i]}(t)
\label{eqn:ErrorDyn}
\end{equation}
 
The adaptive law is determined from Lyapunov's second stability method. A quadratic Lyapunov candidate is defined as a function in terms of $\tilde{{x}}_{[i]}(t)$ and $\tilde{{\theta}}_{[i]}(t)$.
\begin{equation}
\mathcal{V}_{[i]}(\tilde{x}_{[i]}(t), \tilde{\theta}_{[i]}(t)) = \tilde{x}_{[i]}(t)^TP_{i}\tilde{x}_{[i]}(t) + \tilde{\theta}_{[i]}(t)^T\Gamma_{i}^{-1}\tilde{\theta}_{[i]}(t)
\label{eqn:LyapVi}
\end{equation}
where, $P_{i} \in \mathbb{R}^{3\times3}$ is a symmetric matrix, such that $P_{i} = P_{i}^T  > 0$ is the solution to the algebraic Lyapunov linear inequality $A_{ii}^TP_{i} + P_{i}A_{ii} \leq -Q_{i}$, for arbitrary $Q_{i} = Q_{i}^T > 0$, and $\Gamma_{i} \in \mathbb{R}^+$ is the adaptive gain. If the time-derivative of (\ref{eqn:LyapVi}) is at least negative semi-definite, then (\ref{eq:L1SS2}) is locally stable since the energy along the trajectories of state and estimation errors decreases. By defining the adaptive law as,
\begin{equation}
\dot{\hat{\theta}}_{[i]}(t) \triangleq 
\Gamma_i\textrm{Proj}(\hat{\theta}_{[i]}(t), -\bar{x}_{[i]}(t)\tilde{x}_{[i]}^T(t)P_{i}\hat{B}_{m})
\label{eqn:adaptiveLaw1}
\end{equation}
where the projection operator, defined in  \cite{Lavretsky2011}, bounds the parametric uncertainty estimate, then
\begin{equation}
\mathcal{\dot{V}}_{[i]}(\tilde{x}_{[i]}(t), \tilde{\theta}_{[i]}(t))  \leq -\tilde{x}_{[i]}(t)^TQ_{i}\tilde{x}_{[i]}(t) \leq 0
\label{eqn:dotVi3}
\end{equation}

By invoking Barbalat's Lemma it follows that $\lim\limits_{t \rightarrow \infty} \tilde{x}_{[i]}(t) = 0$. Hence, asymptotic convergence of $\tilde{x}_{[i]}(t)$ and boundedness of $\tilde{\theta}_{[i]}(t)$ is proven. 

\subsection{Filter design}
The phase-lag introduced by LPFs can reduce phase-margin and lead to instability. This section defines the conditions for which local stability is ensured when the LPF is inserted. The closed-loop dynamics are presented in the Laplace domain as,
\begin{equation}
\begin{aligned}
\bar{x}_{[i]}(s) =  {(s\mathbb{I}-\hat{A}_{m})}^{-1}\bar{B}_{i}\left(1-C(s)\right)\eta_{[i]}(s)+{(s\mathbb{I}-\hat{A}_{m})}^{-1}Fy_{[i]}^{\textrm{ref}}(s)+{(s\mathbb{I}-\hat{A}_{m})}^{-1}\bar{x}_{[i]}^0
\label{eqn:clref1}
\end{aligned}
\end{equation}
where $\bar{x}_{[i]}^0 = \bar{x}_{[i]}(0)$ is the initial state and,  
\begin{equation}
\eta_{[i]}(t)=\hat{\theta}_{[i]}^T(t)\left(\bar{x}_{[i]}(t)+\tilde{x}_{[i]}(t)\right)
\label{eqn:neta}
\end{equation}
From (\ref{eqn:adaptiveLaw1}) and (\ref{eqn:dotVi3}), the signals $\hat{\theta}_{[i]}(t)$ and $\tilde{x}_{[i]}(t)$ are bounded. 

However, $\bar{\Sigma}_{[i]}^{\textrm{DGU}}$ is non-linear due to the term in (\ref{eqn:neta}). To determine stability conditions upon insertion of the LPF and to prescribe performance specifications such as rise-time, settling-time etc., a linear time-invariant (LTI) reference model is defined where the uncertainty term is known. The available reference system can be given as,
\begin{equation}
\dot{\bar{x}}_{[i]}^{\textrm{ref}}(t) =  \hat{A}_{m}\bar{x}_{[i]}^{\textrm{ref}}(t)+\bar{B}_{i}\left(1-C(p)\right)\bar{\theta}_{[i]}^{*^{T}}\bar{x}_{[i]}^{\textrm{ref}}(t)+Fy_{[i]}^{\textrm{ref}}(t)
\label{eqn:clref2}
\end{equation}
where, $\bar{\theta}_{[i]}(t)=\bar{\theta}_{[i]}^*$ is known. From (\ref{eqn:clref2}), estimation is decoupled from control as the identification of the local state vector is independent of the control input. The performance and robustness specifications can now be set independent of the estimation process. Subsequently, section \ref{DesignConsiderations} shows that the $\bar{\Sigma}_{[i]}^{\textrm{DGU}}$ converges to the LTI reference model with uniform and decoupled performance bounds as the adaptation increases.

Using the LTI reference model, the conditions for local stability due to insertion of the LPF are determined by the following Lemma: \vspace{3mm} \newline 
\textbf{Lemma 1.} \textit{Following the small-gain theorem and Lemma 2.1.2 in \cite{L12010}, if $||(C(s)-1)(s\mathbb{I}-\hat{A}_{m})^{-1}\bar{B}_{i}||_{\mathcal{L}_1}\theta_{\textrm{max}} < 1$, then the reference system (\ref{eqn:clref2}) is bounded-input-bounded-state stable with respect to initial conditions and reference output.}
 \begin{proof}
  From the definition of the closed-loop reference system (\ref{eqn:clref2}), it follows that,
\begin{equation}
\begin{aligned}
\bar{x}_{[i]}^{\textrm{ref}}(s) =  {(s\mathbb{I}-\hat{A}_{m})}^{-1}\bar{B}_{i}\left((1-C(s))\bar{\theta}_{[i]}^{*^{T}}\bar{x}_{[i]}^{\textrm{ref}}(s)\right)+{(s\mathbb{I}-\hat{A}_{m})}^{-1}Fy_{[i]}^{\textrm{ref}}(s)+{(s\mathbb{I}-\hat{A}_{m})}^{-1}\bar{x}_{[i]}^0
\end{aligned}
\label{eqn:clref3}
\end{equation}
As, $(s\mathbb{I}-\hat{A}_{m})^{-1}$ and $C(s)$ are stable transfer functions, the following bound holds,
\begin{equation}
||\bar{x}_{[i]}^{\textrm{ref}}||_{\mathcal{L}_\infty} \leq \frac{||{(s\mathbb{I}-\hat{A}_{m})}^{-1}\bar{x}_{[i]}^0||_{\mathcal{L}_\infty}+||{(s\mathbb{I}-\hat{A}_{m})}^{-1}Fy_{[i]}^{\textrm{ref}}||_{\mathcal{L}_\infty}}{1-||(C(s)-1){(s\mathbb{I}-\hat{A}_{m})}^{-1}\bar{B}_{i}\bar{\theta}_{[i]}^{*^{T}}||_{\mathcal{L}_1}} 
\label{eqn:L1norm}
\end{equation}
Since $\bar{x}_{[i]}^{\textrm{i.c.}}(t)$ and $y_{[i]}^{\textrm{ref}}(t)$ are uniformly bounded, the reference states are bounded as long as the denominator of (\ref{eqn:L1norm}) does not equal zero. As defined in \cite{L12010}, the worst case adaptation bound is,
\begin{equation}
\theta_{\textrm{max}} = 4\max_{\theta \in \Theta}||\bar{\theta}_{[i]}^*||^2_1
\label{eqn:ThetaMax}
\end{equation} 
where $\bar{\theta}_{[i]}^*$ is determined as the required gain such that the eigenvalues of $\bar{A}_{ii}$ and $\hat{A}_m$ are the same. Here, $\bar{A}_{ii}$ is at its most uncertain, i.e. maximal deviation between plant and desired eigenvalues.
$\theta_{\textrm{max}}$ represents the boundary of projection for estimating the parameters when using the adaptation law (\ref{eqn:adaptiveLaw1}). Finally, for the reference states to remain bounded, the following $\mathcal{L}_1$-norm condition must be satisfied,
\begin{equation}
\lambda = ||(C(s)-1)(s\mathbb{I}-\hat{A}_{m})^{-1}\bar{B}_{i}||_{\mathcal{L}_1}\theta_{\textrm{max}} < 1
\label{eqn:L1normCond}
\end{equation} 
where the degree-of-freedom is $\omega_c$. 
\end{proof}
\vspace{-1.5mm}
Consequently, Barbalat's lemma can be used again to show $\lim\limits_{t \rightarrow \infty} \tilde{x}_{[i]}(t) = 0$, i.e. asymptotic tracking is maintained when inserting the LPF. By designing the LPF bandwidth to minimise (\ref{eqn:L1normCond}), the $\mathcal{L}_1$AC guarantees uniform transient and steady-state performance bounds. \vspace{3mm} \newline 
\textbf{Remark 2.} \textit{In order to neglect the high-frequency component of the CPL, the bandwidth of $C(s)$ should be smaller than the closed-loop natural frequency of the overall impedance $\mathbf{Z_{in}(s)}$.}
 \vspace{3mm}
  \subsection{Design considerations}\label{DesignConsiderations}
  This section derives: local stability conditions when CPLs augment the dynamics of each DGU; uniform and decoupled performance bounds of input and output signals between the non-linear system (\ref{eq:L1SS2}) and LTI reference model (\ref{eqn:clref2}); and global stability conditions.
  
  The closed-loop desired dynamics matrix for boost converters is defined as, 
  \begin{equation}
  \hat{A}_{m}^{Boost}=
 \left[\begin{array}{ccc}
-\frac{R_{t_{i}}+V_{dc_{i}}K_i^i}{L_{t_{i}}} & -\frac{(1-D_i+V_{dc_{i}}K_i^v)}{L_{t_{i}}} & -\frac{V_{dc_{i}}K_i^\xi}{L_{t_{i}}}\\
\frac{1-D_i+I_{t_{i}}K_i^i}{C_{t_{i}}} & -\frac{1}{C_{t_{i}}}\left(\sum_{j\in\mathcal{N}_i}\frac{1}{R_{ij}}-\frac{P_{i}^{\textrm{CPL}}}{V_{dc_{i}}^2}-I_{t_{i}}K_i^v\right) & \frac{I_{t_{i}}K_i^\xi}{C_{t_{i}}} \\
 0 & -1 & 0
\end{array} \right] 
\end{equation}

The closed-loop desired dynamics matrix for buck converters is defined as,                                                         \begin{equation}                                                    \hat{A}_{m}^{Buck}=
\left[ \begin{array}{ccc}
-\frac{R_{t_{i}}+K_i^i}{L_{t_{i}}} & -\frac{(1+K_i^v)}{L_{t_{i}}} & 0\\
\frac{1}{C_{t_{i}}} & -\frac{1}{C_{t_{i}}}\left(\sum_{j\in\mathcal{N}_i}\frac{1}{R_{ij}}-\frac{P_{i}^{\textrm{CPL}}}{V_{dc_{i}}^2}\right) & \frac{K_i^\xi}{C_{t_{i}}} \\
0 & -1 & 0
\end{array} \right] 
\label{eqn:Am}                                            
\end{equation}
  Necessary and sufficient conditions for the stability of the local closed-loop subsystems are: \newline $\textrm{trace}(\hat{A}_{m}^{Boost/Buck}) < 0$ and $\textrm{det}(\hat{A}_{m}^{Boost/Buck}) > 0$. As a result, the control gains must satisfy the following conditions in order to preserve the local stability of the DC mG under CPLs. The conditions for boost converters are,
  \begin{equation}
  \begin{aligned}
  K_i^i < -\left(\frac{I_{t_{i}}R_{t_{i}}+V_{dc_{i}}(1-D_i)}{2I_{t_{i}}V_{dc_{i}}}\right)\\
  K_i^v < \frac{1}{I_{t_{i}}}\left(\sum_{j\in\mathcal{N}_i}\frac{1}{R_{ij}}-\frac{P_{i}^{\textrm{CPL}}}{V_{dc_{i}}^2} - \frac{C_{t_{i}}}{L_{t_{i}}}(R_{t_{i}}+K_i^iV_{dc_{i}})\right) \\
  K_i^\xi > 0
  \label{eqn:boostcond}
  \end{aligned}
  \end{equation}
  The conditions for buck converters are,
  \begin{equation}
  \begin{aligned}
  K_i^i < \frac{1}{C_{t_{i}}}\left(\sum_{j\in\mathcal{N}_i}\frac{1}{R_{ij}}-\frac{P_{i}^{\textrm{CPL}}}{V_{dc_{i}}^2}\right) - \frac{R_{t_{i}}}{L_{t_{i}}} \\
  K_i^v < - (R_{t_{i}}+K_i^i)\left(\sum_{j\in\mathcal{N}_i}\frac{1}{R_{ij}}-\frac{P_{i}^{\textrm{CPL}}}{V_{dc_{i}}^2}\right) -1 \\
  K_i^\xi > 0
   \label{eqn:buckcond}
  \end{aligned}
  \end{equation}
  
   The convergence of the non-linear closed-loop adaptive system to the uncertainty free linear reference model in (\ref{eqn:clref2}) can be shown by subtracting (\ref{eqn:clref2}) from the uncertain plant in \label{eqn:clref1}. The bound on this is represented as
  \begin{equation}
      ||\bar{x}_{[i]} - \hat{x}_{[i]}^{\textrm{ref}}||_{\mathcal{L}_{\infty}} \leq \frac{||C(s)||_{\mathcal{L}_1}||\rho_{[i]}(s)||_{\mathcal{L}_1}}{1-||(C(s)-1)(s\mathbb{I}-\hat{A}_{m})^{-1}\bar{B}_{i}||_{\mathcal{L}_1}\theta_{\textrm{max}}}
      \label{x_xref}
  \end{equation}
  where, from Lemma 2.2.5 in \cite{L12010}, $\rho_{[i]}(s)$ is the Laplace transform of the upper-bound on the non-zero state-error initialisation,
  \begin{equation}
  \rho_{[i]}(t)\triangleq\sqrt{\frac{(\mathcal{V}_{[i]}(0)-\frac{\theta_{\textrm{max}}}{\Gamma_{i}})e^{-\alpha_i t}}{\lambda_{\textrm{min}}(P_{i})}+\frac{\theta_{\textrm{max}}}{\lambda_{\textrm{min}}(P_{i})\Gamma_{i}}}\hspace{5mm};\hspace{5mm}\alpha_i \triangleq \frac{\lambda_{\textrm{min}}(Q_{i})}{\lambda_{\textrm{max}}(P_{i})}
  \end{equation}
  (\ref{x_xref}) is rewritten as,
  \begin{equation}
  ||\bar{x}_{[i]} - \hat{x}_{[i]}^{\textrm{ref}}||_{\mathcal{L}_{\infty}} \leq \frac{\gamma_1}{\sqrt{\Gamma_i}}
  \end{equation}
  where,
  $$
  \gamma_1 = \frac{||C(s)||_{\mathcal{L}_1}||\rho_{[i]}(s)||_{\mathcal{L}_1}}{1-||(C(s)-1)(s\mathbb{I}-\hat{A}_{m})^{-1}\bar{B}_{i}||_{\mathcal{L}_1}\theta_{\textrm{max}}}\sqrt{\frac{(\Gamma_i \mathcal{V}_{[i]}(0)-\theta_{\textrm{max}})e^{-\alpha_i t}}{\lambda_{\textrm{min}}(P_{i})}+\frac{\theta_{\textrm{max}}}{\lambda_{\textrm{min}}(P_{i})}}
  $$
 
  For bound between real and LTI reference control inputs, we define,
  \begin{equation}
  \begin{aligned}
      u_{[i]}(s) - u_{[i]}^{\textrm{ref}}(s)= -C(s)\left(\eta_{[i]}(s)+\tilde{\eta}_{[i]}(s)\right)+C(s)\left( \bar{\theta}_{[i]}^{*^{T}}\hat{x}_{[i]}^{\textrm{ref}_(s)}\right)\\
      -C(s)\tilde{\eta}_{[i]}(s)\bar{\theta}_{[i]}^{*^{T}}\left(\bar{x}_{[i]}(s)-\hat{x}_{[i]}^{\textrm{ref}}(s)\right)
      \label{eqn:u_uref}
      \end{aligned}
  \end{equation}
  where, following (\ref{eqn:neta}), $\tilde{\eta}_{[i]}(s)$ is the Laplace transform of $\tilde{\eta}_{[i]}(t) = \hat{\theta}_{[i]^T(t}\tilde{x}_{[i]}(t)$. Using Lemma A.12.1 of \cite{L12010}, which implies that there exists an arbitrary $c_{0}$, we define the strictly proper and bounded-input-bounded-output stable transfer function,
  \begin{equation}
      H_1(s) = C(s)\frac{1}{c_{0}^{T}(s\mathbb{I}-\hat{A}_{m})^{-1}\bar{B}_{i}}c_{0}^{T}
      \label{eqn:H1}
  \end{equation}
  The error-dynamics in (\ref{eqn:ErrorDyn}) can be written in the Laplace domain as,
  \begin{equation}
      \tilde{x}_{[i]}(s) = (s\mathbb{I}-\hat{A}_{m})^{-1}\bar{B}_{i}\tilde{\eta}_{[i]}(s)
      \label{eqn:errorDynLap}
  \end{equation}
  Using (\ref{eqn:H1}) and (\ref{eqn:errorDynLap}), $C(s)\tilde{\eta}_{[i]}(s) = H_1(s)\tilde{x}_{[i]}(s)$, (\ref{eqn:u_uref}) becomes,
  \begin{equation}
      u_{[i]}(s) - u_{[i]}^{\textrm{ref}}(s)= -H_1(s)\tilde{x}_{[i]}(s)+C(s)\bar{\theta}_{[i]}^{*^{T}}\left(\bar{x}_{[i]}(s)-\hat{x}_{[i]}^{\textrm{ref}}(s)\right)
      \label{eqn:u_uref2}
  \end{equation}
 The performance bound of (\ref{eqn:u_uref2}) is written as,
 \begin{equation}
     ||u_{[i]}-u_{[i]}^{\textrm{ref}}|| \leq ||H_1(s)||_{\mathcal{L}_1}||\tilde{x}_{[i]}||_{\mathcal{L}_\infty}+||C(s)||_{\mathcal{L}_1}\theta_{\textrm{max}}\gamma_1
 \end{equation}
 or,
 \begin{equation}
     ||u_{[i]}-u_{[i]}^{\textrm{ref}}|| \leq \frac{\gamma_2}{\sqrt{\Gamma_i}} 
 \end{equation}
 where,
 \begin{equation}
 \gamma_2 = || C(s)\frac{1}{c_{0}^{T}(s\mathbb{I}-\hat{A}_{m})^{-1}\bar{B}_{i}}c_{0}^{T}||_{\mathcal{L}_1}\sqrt{\frac{(\Gamma_i \mathcal{V}_{[i]}(0)-\theta_{\textrm{max}})e^{-\alpha_i t}}{\lambda_{\textrm{min}}(P_{i})}+\frac{\theta_{\textrm{max}}}{\lambda_{\textrm{min}}(P_{i})}}+||C(s)||_{\mathcal{L}_1}\theta_{\textrm{max}}\gamma_1
 \label{eqn:Gamma2}
  \end{equation} \vspace{3mm} \newline 
  \textbf{Remark 3.} \textit{It is clear that if the relative degree of $C(s)$ is not at least equal to the relative degree of the plant, $(s\mathbb{I}-\hat{A}_{m})^{-1}\bar{B}_{i}$, then the first term in (\ref{eqn:Gamma2}) is not strictly proper, and will result in unbounded control inputs, and therefore no transient guarantees can be provided. This is also why in standard MRAC architectures, i.e. $C(s)=1$, such guarantees cannot be provided.}\vspace{3mm} \newline \textbf{Remark 4.} \textit{The inversion of $(s\mathbb{I}-\hat{A}_{m})^{-1}\bar{B}_{i}$ results in unstable poles for non-minimum-phase systems. Boost converters are non-minimum-phase for duty-cycle to output voltage control. Therefore, the desired dynamics must be carefully designed to be minimum-phased. This can be done by transforming local systems to control-canonical form, as performed in \cite{OKeeffe2018c}.}
  \vspace{3mm}
  Finally, $\lim\limits_{\Gamma_i\rightarrow \infty}||\bar{x}_{[i]} - \hat{x}_{[i]}^{\textrm{ref}}||_{\mathcal{L}_{\infty}} = 0$ and $\lim\limits_{\Gamma_i\rightarrow \infty}||u_{[i]} - u_{[i]}^{\textrm{ref}}||_{\mathcal{L}_{\infty}} = 0$ shows that the non-linear local plant converges to the linear reference model, and the performance bounds of both state and control inputs decrease as the adaptive gain increases.
  
  Conservative global asymptotic stability conditions are derived offline using collective Lyapunov functions. The overall Lyapunov function candidate that describes the global system can be written as,
\begin{equation}
 \mathcal{V}(t) = \sum\limits_{i=0}^{N} \left(\tilde{x}_{[i]}^T(t)P_{i}\tilde{x}_{[i]}(t) + \tilde{\theta}_{[i]}^T(t)\Gamma_{i}^{-1}\tilde{\theta}_{[i]}(t)\right)
 \label{eqn:OverallLyap}
 \end{equation}
 \textbf{Assumption 1.} \textit{We assume local controllers exploit (\ref{eqn:adaptiveLaw1}), and  plant dynamics have converged to desired dynamics.}
 \vspace{3mm} \newline
The derivative of (\ref{eqn:OverallLyap}) is,
 \begin{equation}
  \dot{\mathcal{V}} = -\sum\limits_{i=0}^{N}\tilde{x}_{[i]}^T(t)Q_{i}\tilde{x}_{[i]}(t)
  \label{eq:OverallVdot}
 \end{equation}
 if and only if matrix $\mathbf{P}$ satisfies the Lyapunov inequality equation,
 \begin{equation} \underbrace{\hat{\textbf{A}}_{\textbf{m}}^T\textbf{P}+\textbf{P}\hat{\textbf{A}}_{\textbf{m}}}_{(\textrm{a})}+\underbrace{\hat{\textbf{A}}_{\textbf{C}}^T\textbf{P}+\textbf{P}\hat{\textbf{A}}_{\textbf{C}}}_{(\textrm{b})} < 0
  \label{eq:OverallVdot2}
 \end{equation} 
 where $\textbf{P} = $ diag$({P}_{i}) \in \mathbb{R}^{3M\times 3M}$; $\hat{\textbf{A}}_{\textbf{m}} =$ diag$(\hat{A}_{m}) \in \mathbb{R}^{3M\times3M}$ represents the overall desired dynamics; $\hat{\textbf{A}}_{\textbf{C}} = \hat{\textbf{A}} - \hat{\textbf{A}}_{\textbf{m}} \in \mathbb{R}^{3M\times3M}$ represents the coupling dynamics only. As each DGU is designed to be locally asymptotically stable, the matrices of (a) are negative definite. Therefore, the design of $\textbf{K}$ is performed iteratively offline to ensure $||(a)||$ $>$ $||(b)||$. This typically results in detuned controller gains, as expected due to the conservative requirements of decentralised systems. Furthermore, this method can suffer when system size expands as the retuning of $\textbf{K}$ becomes more difficult, despite the advantage of designing the desired dynamics as the same for all DGUs. The controller gains can be tuned optimally using LQR guidelines, such as in \cite{Kurucs2015}, provided conditions (\ref{eqn:boostcond}) or (\ref{eqn:buckcond}) depending on converter topology, and (\ref{eq:OverallVdot2}) are satisfied.

\begin{figure}[!htb]    
  \graphicspath{{Images/}}
  \centering
  \includegraphics[width=13cm]{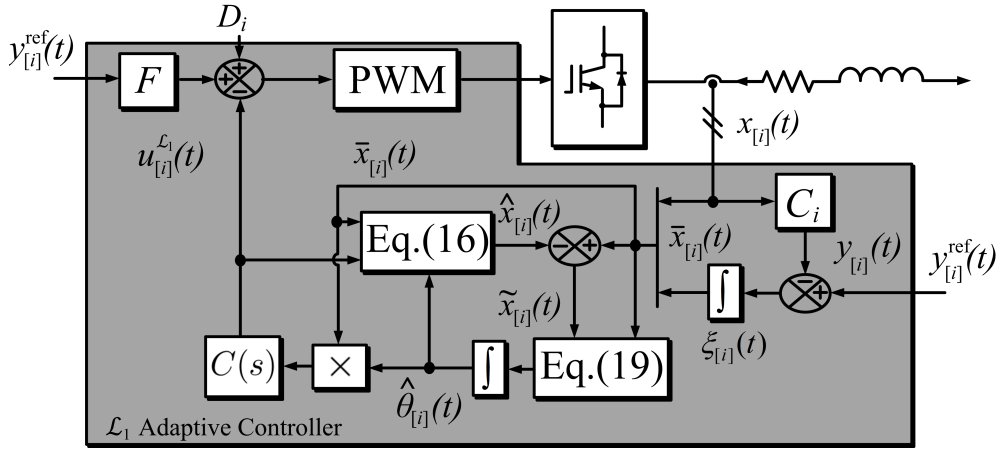}
  \caption{Purely adaptive architecture: decentralised $\mathcal{L}_1$ adaptive controller.}
  \label{fig:OverallArch}
  \end{figure}

  The decentralised $\mathcal{L}_1$ adaptive primary control architecture can be seen in Fig.\ref{fig:OverallArch}.
  \section{Coordinated secondary control}
  Coordination among multiple DGUs within any mG is imperative for dynamic operation and system management. Two key control objectives include: (i) voltage restoration, and (ii) load-power sharing \cite{Vasquez2016,Lu2014}. In reference to (i), as the bus voltage of a bus-connected mG cannot be directly controlled and since unknown voltage drops occur across unknown power-lines, the secondary control level is required to compute appropriate voltage references for the primary control level in order to maintain each DGU voltage within a prescribed range. In reference to (ii), secondary controllers are required to also provide a voltage reference that ensures DGUs can share the power delivery between each other independent of both topology and power-lines. The mathematical definitions are as follows.\vspace{3mm} \newline 
  \textbf{Definition 1.} \textit{Equal-current sharing is achieved if the output current of each DGU equals the average load-current, i.e. }
\begin{equation}
      i_{i}^{\textrm{out}} = \langle \textbf{I}_{\textbf{L}} \rangle
      \label{eqn:shareI}
\end{equation}
  \textit{where,} $i_{i}^{\textrm{out}} = (1-D_i)i_{dc_{i}}$ \textit{for boost converters}, $i_{i}^{\textrm{out}} = i_{dc_{i}}$ \textit{for buck converters}, $\textbf{I}_{\textbf{L}} = [i_{L_{1}},i_{L_{2}},...,i_{L_{N}}]^T$ \textit{is the local unknown load current vector, and the operator} $\langle . \rangle$ \textit{denotes the average value of a vector.}\vspace{3mm} \newline 
  \textbf{Assumption 2.} \textit{Secondary control voltage references are identical} $V_{\textrm{bus}_{i}}^{\textrm{ref}}$ = $V_{\textrm{bus}}^{\textrm{ref}}$.\vspace{3mm} \newline 
  \textbf{Definition 2.} \textit{Under assumption 2, voltage restoration is achieved if the average voltage of all DGU output voltages equals} $V_{\textrm{bus}}^{\textrm{ref}}$, \textit{i.e.}
\begin{equation}
      \langle \textbf{\textrm{V}} \rangle = V_{\textrm{bus}}^{\textrm{ref}}
      \label{eqn:av}
\end{equation}
  \textit{where}, $\textbf{V} = [v_{dc_{1}},v_{dc_{2}},...,v_{dc_{N}}]^T$.
  \vspace{3mm} \newline
  \textbf{Remark 5.} \textit{We consider a network of communication links consisting of a set of nodes} $\mathcal{T}=\{T_1,T_2,...,T_N\}$, \textit{connected through edges} $\mathcal{E}=\mathcal{T}\times\mathcal{T}$. \textit{Such a network is described by a graph} $\mathcal{G}=(\mathcal{T}, \mathcal{E})$. \textit{Each node represents a DGU in the network and edges represent communication links for information exchange. The Laplacian matrix,} $\mathbb{L}(\mathcal{G})$ \textit{describes the graph in matrix form, and is defined as,}
  \begin{equation}
  \mathbb{L}(\mathcal{G})_{ij}=
  \begin{cases}
  -1, \hspace{9.5mm}\textrm{if} \hspace{1.5mm} i\neq j \hspace{1.5mm}\textrm{and}\hspace{1.5mm} T_i \hspace{1.5mm}\textrm{is adjacent to}\hspace{1.5mm} T_j \\
  \textrm{deg}(T_i), \hspace{3mm}\textrm{if} \hspace{1.5mm}i = j \\
  0, \hspace{12.5mm}\textrm{otherwise}
  \end{cases}
  \end{equation}
  \textit{where} deg($T_i$) \textit{represents the number of adjacent nodes to connected to} $T_i$.
  
  \vspace{3mm}
  The secondary control level is typically implemented at a slower bandwidth to the primary level, and utilises a communication network in order to coordinate and manage voltage and current/power levels within the overall mG. 
  
  \subsection{Distributed consensus-based controller design}
  In a distributed system, the bus voltage of a bus-connected mG cannot be directly controlled. Instead, each primary level voltage reference is controlled such that the average of all the local voltages equals an estimate of the bus. Secondary controllers in \cite{Andreasson2016,Riverso2017a,OKeeffe2017a,Lu2014} have utilised a fully-connected low-bandwidth communications (LBC) network, whereby the average voltage estimate is determined by measuring every DGU. However, this clearly places restrictions on the network i.e. fault-tolerance. As a result, a sparse-connected, peer-to-peer, LBC network has been proposed, whereby dynamic consensus algorithms are implemented to estimate global information from a limited number of nodes \cite{Nasirian2014a,Shafiee2014b}.
  
  The dynamic consensus algorithm used to estimate the mG bus voltage is denoted as,
  \begin{subequations}
  \begin{align}
      \hat{v}_{i}^{\textrm{bus}}(t) \triangleq v_{dc_{i}}(t)+\int\limits_{0}^{t}\sum\limits_{j\in\mathcal{N}_i}a_{ij}\left(\hat{v}_{j}^{\textrm{bus}}(\tau)-\hat{v}_{i}^{\textrm{bus}}(\tau)\right)d\tau = \\
      v_{dc_{i}}(t)-\int\limits_{0}^{t}\sum\limits_{j\in\mathcal{N}_i}a_{ij}\left(\hat{v}_{i}^{\textrm{bus}}(\tau)-\hat{v}_{j}^{\textrm{bus}}(\tau)\right)d\tau
      \end{align}
  \end{subequations}
  where $a_{ij}=1$ if $ \bar{\Sigma}_{[j]}^{\textrm{DGU}}$ is connected to $\bar{\Sigma}_{[i]}^{\textrm{DGU}}$, otherwise $a_{ij}=0$. Subsequently, the estimate of the mG bus voltage is used to drive the average of all the voltages to equal a bus voltage reference.
 
  To achieve (ii), the voltage reference for the primary level is controlled to ensure equal current sharing, i.e. the current injected equals the average total current injected within the mG. A consensus algorithm similar to above is also used for this.
  \begin{subequations}
  \begin{align}
      i_{\textrm{ref}_{i}}^{\textrm{out}}(t) \triangleq i_{i}^{\textrm{out}}(t)+\int\limits_{0}^{t}\sum\limits_{j\in\mathcal{N}_i}a_{ij}\left(\hat{i}_{j}^{\textrm{out}}(\tau)-\hat{i}_{i}^{\textrm{out}}(\tau)\right)d\tau = \\
      i_{i}^{\textrm{out}}(t)-\int\limits_{0}^{t}\sum\limits_{j\in\mathcal{N}_i}a_{ij}\left(\hat{i}_{i}^{\textrm{out}}(\tau)-\hat{i}_{j}^{\textrm{out}}(\tau)\right)d\tau
      \label{eqn:avIout}
      \end{align}
  \end{subequations}
Using remark 5, the global consensus algorithm can be represented by vector form as,
\begin{subequations}
\begin{align}
  \hat{\textbf{V}}_{\textrm{bus}} \triangleq \textbf{V} - \mathbb{L}(\mathcal{G})\int \hat{\textbf{V}}d\tau =\\
  \textbf{V} + \mathbb{L}(\mathcal{G})\int \hat{\textbf{V}}d\tau
  \end{align}
\end{subequations}
depending on how the vector $\hat{\textbf{V}}$ is constructed. Similarly,
\begin{subequations}
\begin{align}
    \textbf{I}_{\textrm{ref}} = \textbf{I} - \mathbb{L}(\mathcal{G})\int \hat{\textbf{I}}d\tau =\\
  \textbf{I} + \mathbb{L}(\mathcal{G})\int \hat{\textbf{I}}d\tau
  \end{align}
  \label{eqn:IconvGlobal}
\end{subequations}
depending on how the vector $\hat{\textbf{I}}$ is constructed.

  Finally, the following controlled correction terms can be added to $V_{\textrm{bus}}^{\textrm{ref}}$,
  \begin{equation}
      \Delta v_{dc_{i}}(t) = k_{\textrm{P}_{i}}^{v}\left(v_{\textrm{bus}}^{\textrm{ref}}-\hat{v}_{i}^{\textrm{bus}}(t)\right)+k_{\textrm{I}_{i}}^{v}\int\limits_{0}^{t}\left(v_{\textrm{bus}}^{\textrm{ref}}-\hat{v}_{i}^{\textrm{bus}}(\tau)\right)d\tau
      \end{equation}
   \begin{equation}
        \Delta i_{i}^{\textrm{out}}(t) = k_{\textrm{P}_{i}}^{i}\left(i_{\textrm{ref}_{i}}^{\textrm{out}}-\frac{i_{i}^{\textrm{out}}(t)}{m_i}\right)+k_{\textrm{I}_{i}}^{i}\int\limits_{0}^{t}\left(i_{\textrm{ref}_{i}}^{\textrm{out}}-\frac{i_{i}^{\textrm{out}}(\tau)}{m_i}\right)d\tau
        \end{equation}
        where $m_i$ is the load-sharing co-efficient i.e. for equal current-sharing $m_i = m_j$, and $\Delta v_{dc_{i}}$ and $\Delta i_{i}^{\textrm{out}}$ are generated from PI controllers, represented in the Laplace domain as,
        \begin{equation}
            C_{v}(s) = k_{\textrm{P}_{i}}^{v}+\frac{k_{\textrm{I}_{i}}^{v}}{s} 
             \end{equation}
             and,
             \begin{equation}
            C_{i}(s) = k_{\textrm{P}_{i}}^{i}+\frac{k_{\textrm{I}_{i}}^{i}}{s}
        \end{equation}
        The PI controllers can be tuned by representing the primary level as a unit-gain approximation \cite{Tucci2017g,Han2017}. Finally, the voltage reference sent to the proposed primary control level to achieve (i) and (ii) is given as,
  \begin{equation}
    V_{dc_{i}}^{\textrm{ref}} = v_{\textrm{bus}}^{\textrm{ref}} + \Delta v_{dc_{i}}(t) + \Delta i_{i}^{\textrm{out}}(t)
    \end{equation}
   
\begin{figure}[!htb]    
  \graphicspath{ {Images/} }
  \centering
  \includegraphics[width=13cm]{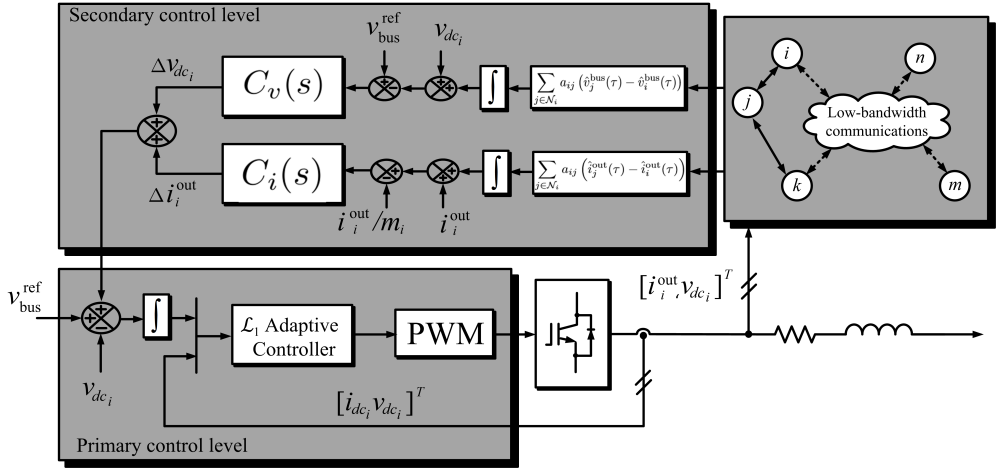}
  \caption{Overall hierarchical control architecture with decentralised $\mathcal{L}_1$ adaptive primary controller and distributed secondary PI controllers.}
  \label{fig:OverallArch2}
  \end{figure}

The overall distributed hierarchical control structure can be seen Fig.\ref{fig:OverallArch2},
  
  \subsection{Stability analysis}
  This section uses Lyapunov stability analysis to show that under the effect of secondary control, stability of the overall system is maintained and all DGU output currents and voltages converge such that (\ref{eqn:shareI}) and (\ref{eqn:av}) are satisfied. Following \cite{Tucci2016}, unit-gain approximations of the primary $\mathcal{L}_1$AC level are made and subsequently, stability is proven.  
  
  \subsubsection{Unit-gain approximation}
  Approximating the primary $\mathcal{L}_1$ adaptive control loops with ideal unitary gains is a reasonable one at low-frequencies since the loop can be shaped by the LPF of $\mathcal{L}_1$AC. The global relationship between the primary $\mathcal{L}_1$AC level and the distributed-consensus based secondary control level is represented as,
  \begin{equation}
      \textbf{V} = \textbf{V}_{\textrm{bus}}^{\textrm{ref}} + \Delta\textbf{V} + \Delta\textbf{I}
      \label{eqn:globalDyn}
  \end{equation}
  where $\Delta\textbf{V} = [\Delta v_{dc_{1}},...,\Delta v_{dc_{2}},...,\Delta v_{dc_{N}}]^T$ and $\Delta\textbf{I} = [\Delta i_{1}^{\textrm{out}},...,\Delta i_{2}^{\textrm{out}},...,\Delta i_{N}^{\textrm{out}}]^T$. The overall adjustment signals from the PI controllers are defined in vector form as,
  \begin{equation}
      \Delta\textbf{V} = \textbf{K}_{\textbf{P}}^{\textbf{V}}\textbf{e}_{\textbf{V}} +  \textbf{K}_{\textbf{I}}^{\textbf{V}} \int \textbf{e}_{\textbf{V}} d\tau
      \label{eqn:globalPIv1}
  \end{equation}
  and, 
  \begin{equation}
      \Delta\textbf{I} = \textbf{K}_{\textbf{P}}^{\textbf{I}}\textbf{e}_{\textbf{I}} +  \textbf{K}_{\textbf{I}}^{\textbf{I}}\int \textbf{e}_{\textbf{I}} d\tau
      \label{eqn:globalPIi1}
  \end{equation}
  where $\textbf{K}_{\textbf{P}}^{\textbf{V}} = \textrm{diag}(k_{\textrm{P}_{i}}^{v})$, $\textbf{K}_{\textbf{I}}^{\textbf{V}} = \textrm{diag}(k_{\textrm{I}_{i}}^{v})$, $\textbf{K}_{\textbf{P}}^{\textbf{I}} = \textrm{diag}(k_{\textrm{P}_{i}}^{i})$, and $\textbf{K}_{\textbf{I}}^{\textbf{I}} = \textrm{diag}(k_{\textrm{I}_{i}}^{i})$. The global secondary control error dynamics are defined as,
  \begin{equation}
      \textbf{e}_{\textbf{V}} = \textbf{V}_{\textrm{bus}}^{\textrm{ref}} - \textbf{V} - \mathbb{L}(\mathcal{G})\int \hat{\textbf{V}}d\tau
      \label{eqn:globalPIv2}
  \end{equation}
  where $\hat{\textbf{V}} = [\hat{v}_{1}^{\textrm{bus}}, \hat{v}_{2}^{\textrm{bus}},...,\hat{v}_{N}^{\textrm{bus}}]^T$, and,
   \begin{equation}
      \textbf{e}_{\textbf{I}} = \textbf{I}_{\textrm{ref}} -  \textbf{I} - \mathbb{L}(\mathcal{G})\int \hat{\textbf{I}}d\tau
      \label{eqn:globalPIi2}
  \end{equation}
   where $\hat{\textbf{I}}= [\hat{i}_{1}^{\textrm{out}}, \hat{i}_{2}^{\textrm{out}},...,\hat{i}_{N}^{\textrm{out}}]^T$.\vspace{3mm} \newline \textbf{Lemma 2.} \textit{The symmetric Laplacian matrix} $\mathbb{L}(\mathcal{G})$ \textit{is positive definite.}\begin{proof}
 An arbitrary vector, $z  \in \mathbb{R}^{n}$, can always be written as z = $\hat{z} + \bar{z}$, where $\hat{z} \in \mathbb{H}^1$ and $\bar{z} \in \mathbb{H}_{\perp}^1$. $\mathbb{H}^1\in \mathbb{R}$ is a subspace composed of vectors with zero average, i.e. $\mathbb{H}^1 = \{\hat{z}\in\mathbb{R}^n:\langle \hat{z} \rangle = 0\}$ and $\mathbb{H}_{\perp}^1\in \mathbb{R}$ is a subspace orthogonal to $\mathbb{H}^1$, i.e. $\mathbb{H}_{\perp}^1 = \{\alpha \textbf{1}_\textbf{n}:\alpha \in \mathbb{R}\}$. Then,
 \begin{equation}
     z^T\mathbb{L}(\mathcal{G})z
 \end{equation}
 is equivalent to the following cases,
 \begin{equation}
     \begin{cases}
     \textrm{If} \hspace{1.5mm} z > 0, \hspace{1.5mm}\textrm{then}  \hspace{1.5mm}z^T\mathbb{L}(\mathcal{G})z > 0 \\
     \textrm{If}\hspace{1.5mm} z < 0,\hspace{1.5mm} \textrm{then} \hspace{1.5mm} z^T\mathbb{L}(\mathcal{G})z > 0 \\
     \textrm{If} \hspace{1.5mm}\bar{z} = 0, \hspace{1.5mm}\textrm{then}\hspace{1.5mm} \hat{z}^T\mathbb{L}(\mathcal{G})\bar{z} > 0
     \end{cases}
 \end{equation}
 Therefore, $\mathbb{L}(\mathcal{G})$ is positive definite.
 \end{proof}
  For convenience, we prove convergence of the output current loop first by isolating the current-sharing adjustment term in (\ref{eqn:globalDyn}), which is expressed as,
  \begin{equation}
      \textbf{I} = \textbf{I}_{\textrm{pri}}^{\textrm{ref}}+\Delta \textbf{I}
      \label{eqn:CLcurrentDyn}
  \end{equation}
  where $\textbf{I}_{\textrm{pri}}^{\textrm{ref}} = \textrm{diag}(\textbf{0}_N)$. The following Lyapunov function candidate is considered,
 \begin{equation}
     \mathcal{V}_\textbf{I} = \frac{1}{2}\textbf{e}_{\textbf{I}}^T \textbf{P}_{\textbf{I}}\textbf{e}_{\textbf{I}}
     \label{eqn:Vi}
 \end{equation}
 The derivative of which equates to,
 \begin{equation}
 \dot{\mathcal{V}}_\textbf{I} \leq \textbf{e}_{\textbf{I}}^T \textbf{P}_{\textbf{I}}\dot{\textbf{e}}_{\textbf{I}}
 \label{eqn:Vidot}
 \end{equation}
 where the derivative of (\ref{eqn:globalPIi2}) is,
  \begin{equation}
 \dot{\textbf{e}}_{\textbf{I}} = - \left(\dot{\textbf{I}} + \mathbb{L}(\mathcal{G})\hat{\textbf{I}}\right)
 \end{equation}
 and the derivative of (\ref{eqn:CLcurrentDyn}) yields,
 \begin{equation}
     \dot{\textbf{I}} = \dot{\Delta \textbf{I}} = -(\mathbb{I}+\textbf{K}_{\textbf{P}}^{\textbf{I}})^{-1}\left(\textbf{K}_{\textbf{P}}^{\textbf{I}}\mathbb{L}(\mathcal{G})\hat{\textbf{I}}-\textbf{K}_{\textbf{I}}^{\textbf{I}}\textbf{e}_{\textbf{I}}\right)
 \end{equation}
Therefore, (\ref{eqn:Vidot}) can be written as,
\begin{equation}
\dot{\mathcal{V}}_\textbf{I} \leq -\textbf{e}_{\textbf{I}}^T\textbf{P}_{\textbf{I}}\left((\mathbb{I}+\textbf{K}_{\textbf{P}}^{\textbf{I}})^{-1}(-\textbf{K}_{\textbf{P}}^{\textbf{I}}\mathbb{L}(\mathcal{G})\hat{\textbf{I}}+\textbf{K}_{\textbf{I}}^{\textbf{I}}\textbf{e}_{\textbf{I}}+\mathbb{L}(\mathcal{G})\hat{\textbf{I}})\right)
\end{equation}
or,
\begin{equation}
\dot{\mathcal{V}}_\textbf{I} \leq -\frac{\textbf{K}_{\textbf{I}}^{\textbf{I}}}{2}\textbf{e}_{\textbf{I}}^T\left(\textbf{P}_{\textbf{I}}\mathcal{A}+\mathcal{A}^T\textbf{P}_{\textbf{I}}\right)\textbf{e}_{\textbf{I}} - \frac{\textbf{K}_{\textbf{P}}^{\textbf{I}}}{2}\textbf{e}_{\textbf{I}}^T\left(\textbf{P}_{\textbf{I}}\mathcal{A}+\mathcal{A}^T\textbf{P}_{\textbf{I}}\right)\mathbb{L}(\mathcal{G})\hat{\textbf{I}} 
\label{eqn:Vidot1}
\end{equation}
where $\mathcal{A}=(\mathbb{I}+\textbf{K}_{\textbf{P}}^{\textbf{I}})^{-1}$. To prove convergence, Barbalat's lemma is invoked. The derivative of (\ref{eqn:Vidot1}) is,
\begin{equation}
\ddot{\mathcal{V}}_\textbf{I} \leq \frac{\textbf{K}_{\textbf{I}}^{\textbf{I}}}{2}\textbf{e}_{\textbf{I}}^T\left(\textbf{P}_{\textbf{I}}\mathcal{A}+\mathcal{A}^T\textbf{P}_{\textbf{I}}\right)\left(\mathcal{A}\textbf{K}_{\textbf{P}}^{\textbf{I}}\mathbb{L}(\mathcal{G})\hat{\textbf{I}}+\mathcal{A}\textbf{K}_{\textbf{I}}^{\textbf{I}}\textbf{e}_{\textbf{I}}\right)
\label{eqn:Viddot}
\end{equation}
Since (\ref{eqn:Vi}) and (\ref{eqn:Viddot}) are bounded, (\ref{eqn:Vidot1}) is uniformly continuous and $\lim\limits_{t\rightarrow \infty}\dot{\mathcal{V}}_{\textbf{I}} = 0$. As both individual terms in $(\ref{eqn:Vidot1})$ are positive definite, both must converge to 0 if $\lim\limits_{t\rightarrow \infty}\dot{\mathcal{V}}_{\textbf{I}} = 0$. As a result, the output currents converge to their reference.\vspace{3mm} \newline 
\textbf{Remark 6.} \textit{A more trivial solution to above can be given if }$\textbf{I}_{\textrm{ref}} = \textbf{I},$ \textit{which is the case when the load-sharing co-efficients} $m_i = m_j = 1$. \textit{As a result} $\textbf{e}_{\textbf{I}} = -\mathbb{L}(\mathcal{G})\int \hat{\textbf{I}}d\tau$, \textit{and therefore,} $\dot{\mathcal{V}}_{\textbf{I}} = -\textbf{e}_{I}^T \textbf{P}_{\textbf{I}}\mathbb{L}(\mathcal{G})\hat{\textbf{I}}$.
\vspace{3mm}

Next, the following Lyapunov function is considered to prove voltage stability and convergence,
\begin{equation}
    \mathcal{V}_{\textbf{V}} = \frac{1}{2}\textbf{e}_{\textbf{V}}^{T}\textbf{P}_{\textbf{V}}\textbf{e}_{\textbf{V}}
\end{equation}
The derivative of which equates to,
\begin{equation}
\dot{\mathcal{V}}_{\textbf{V}} \leq -\textbf{e}_{\textbf{V}}^{T}\textbf{P}_{\textbf{V}}\left(\dot{\textbf{V}}+L\hat{\textbf{V}}\right)
\end{equation}
where, 
\begin{equation}
    \dot{\textbf{V}} = \dot{\Delta \textbf{V}}+\dot{\Delta \textbf{I}}
\end{equation}
Using (\ref{eqn:IconvGlobal}) and similar analysis from above, 

\begin{equation}
\begin{aligned}
    \dot{\mathcal{V}}_{\textbf{V}} \leq - \frac{1}{2}\textbf{K}_{\textbf{I}}^{\textbf{V}}\textbf{e}_{\textbf{V}}^{T}\left(\textbf{P}_{\textbf{V}}\mathcal{A}+\mathcal{A}^T\textbf{P}_{\textbf{V}}\right)\textbf{e}_{\textbf{V}}-\frac{1}{2}\mathbb{L}(\mathcal{G})\textbf{e}_{\textbf{V}}^{T}\left(\textbf{P}_{\textbf{V}}\mathcal{A}+\mathcal{A}^T\textbf{P}_{\textbf{V}}\right)\hat{\textbf{V}}-\\ \frac{1}{2}\textbf{K}_{\textbf{P}}^{\textbf{I}}\mathbb{L}(\mathcal{G})\textbf{e}_{\textbf{V}}^{T}\left(\textbf{P}_{\textbf{V}}\mathcal{A}+\mathcal{A}^T\textbf{P}_{\textbf{V}}\right)\hat{\textbf{I}}-\frac{1}{2}\textbf{e}_{\textbf{V}}^{T}\left(\textbf{P}_{\textbf{V}}\mathcal{A}+\mathcal{A}^T\textbf{P}_{\textbf{V}}\right)\textbf{e}_{\textbf{I}}
    \label{eqn:VdotV}
    \end{aligned}
\end{equation}
To prove asymptotic stability, we again invoke Barbalat's lemma where a bounded second derivative,
\begin{equation}
    \ddot{V}_{\textbf{V}} < \frac{1}{2}\textbf{K}_{\textbf{I}}^{\textbf{V}}\textbf{e}_{\textbf{V}}^{T}\left(\textbf{P}_{\textbf{V}}\mathcal{A}+\mathcal{A}^T\textbf{P}_{\textbf{V}}\right)^2\textbf{e}_{\textbf{V}} + \frac{1}{2}\textbf{K}_{\textbf{I}}^{\textbf{V}}\textbf{e}_{\textbf{V}}^{T}\left(\textbf{P}_{\textbf{V}}\mathcal{A}+\mathcal{A}^T\textbf{P}_{\textbf{V}}\right)^2\hat{\textbf{V}}+\\ \frac{1}{2}\textbf{K}_{\textbf{P}}^{\textbf{I}}\mathbb{L}(\mathcal{G})\textbf{e}_{\textbf{V}}^{T}\left(\textbf{P}_{\textbf{V}}\mathcal{A}+\mathcal{A}^T\textbf{P}_{\textbf{V}}\right)^2\hat{\textbf{I}}+\frac{1}{2}\textbf{e}_{\textbf{V}}^{T}\left(\textbf{P}_{\textbf{V}}\mathcal{A}+\mathcal{A}^T\textbf{P}_{\textbf{V}}\right)^2\textbf{e}_{\textbf{I}}
\end{equation}
leads to a uniformly continuous (\ref{eqn:VdotV}) and $\lim\limits_{t\rightarrow \infty}\dot{\mathcal{V}}_{\textbf{V}} = 0$. Similar to before, the average of each output voltage converges to the bus voltage reference.

  \section{Results}
  Simulations are performed in Matlab/Simulink. Firstly, we show instability induced by negative-incremental impedance between two DC-DC power converters using non-adaptive state-feedback controllers. Subsequently, the proposed architecture is evaluated using a bus-connected mG consisting of 6 DGUs and 2 power electronic loads.
  
  \subsection{Constant-power load instability}
\begin{figure}[!htb]
                 \centering
                 \begin{subfigure}[!htb]{0.52\textwidth}
                   \centering
  \includegraphics[width=\textwidth]{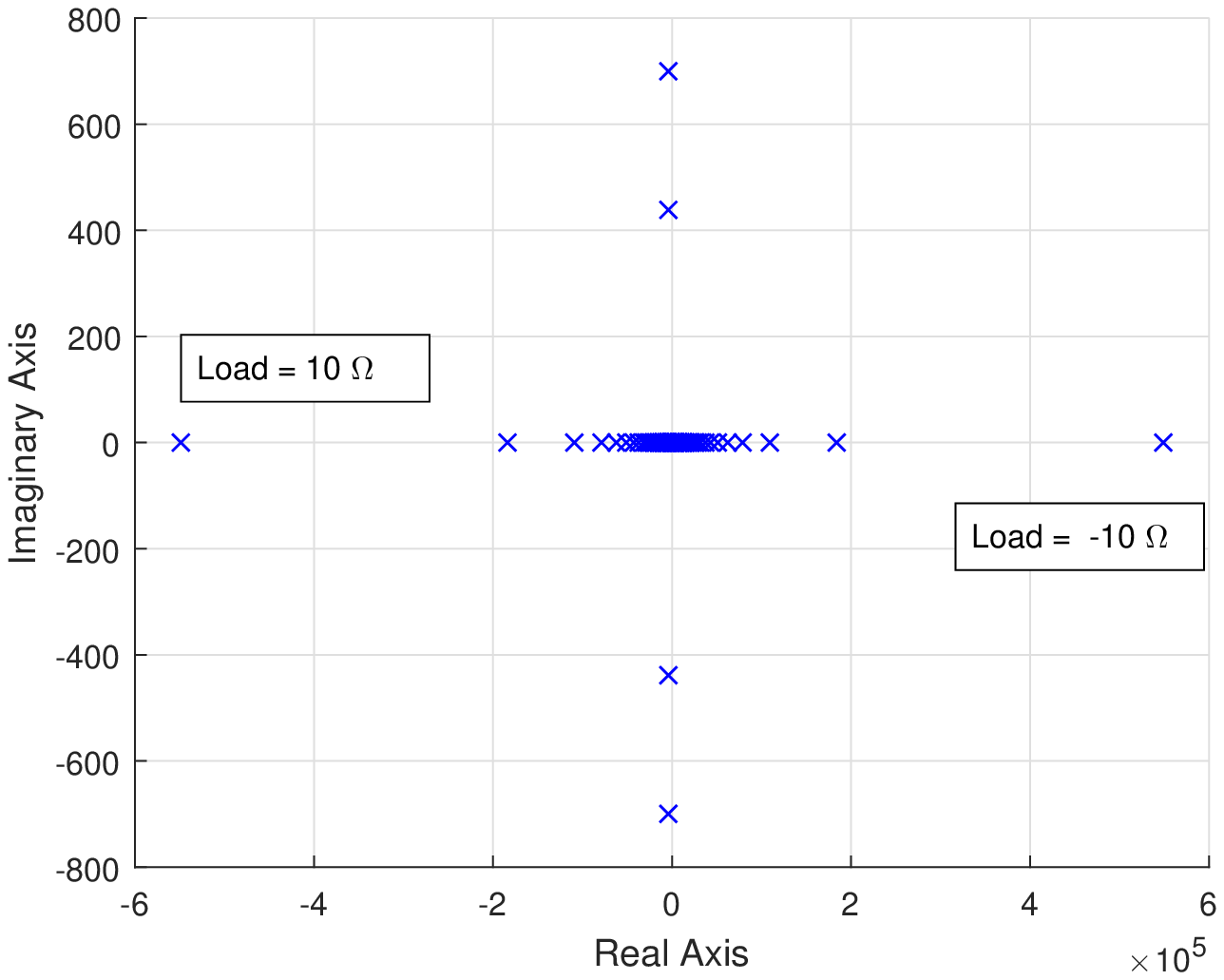}
 \caption{Family of closed-loop eigenvalues as $R_{i}^{\textrm{CPL}}$ varies from -10 $\Omega$ to 10 $\Omega$.}
                   \label{fig:Eig}
                 \end{subfigure}\hspace*{\fill}
                 \begin{subfigure}[!htb]{0.52\textwidth}
                   \centering
 \includegraphics[width=\textwidth]{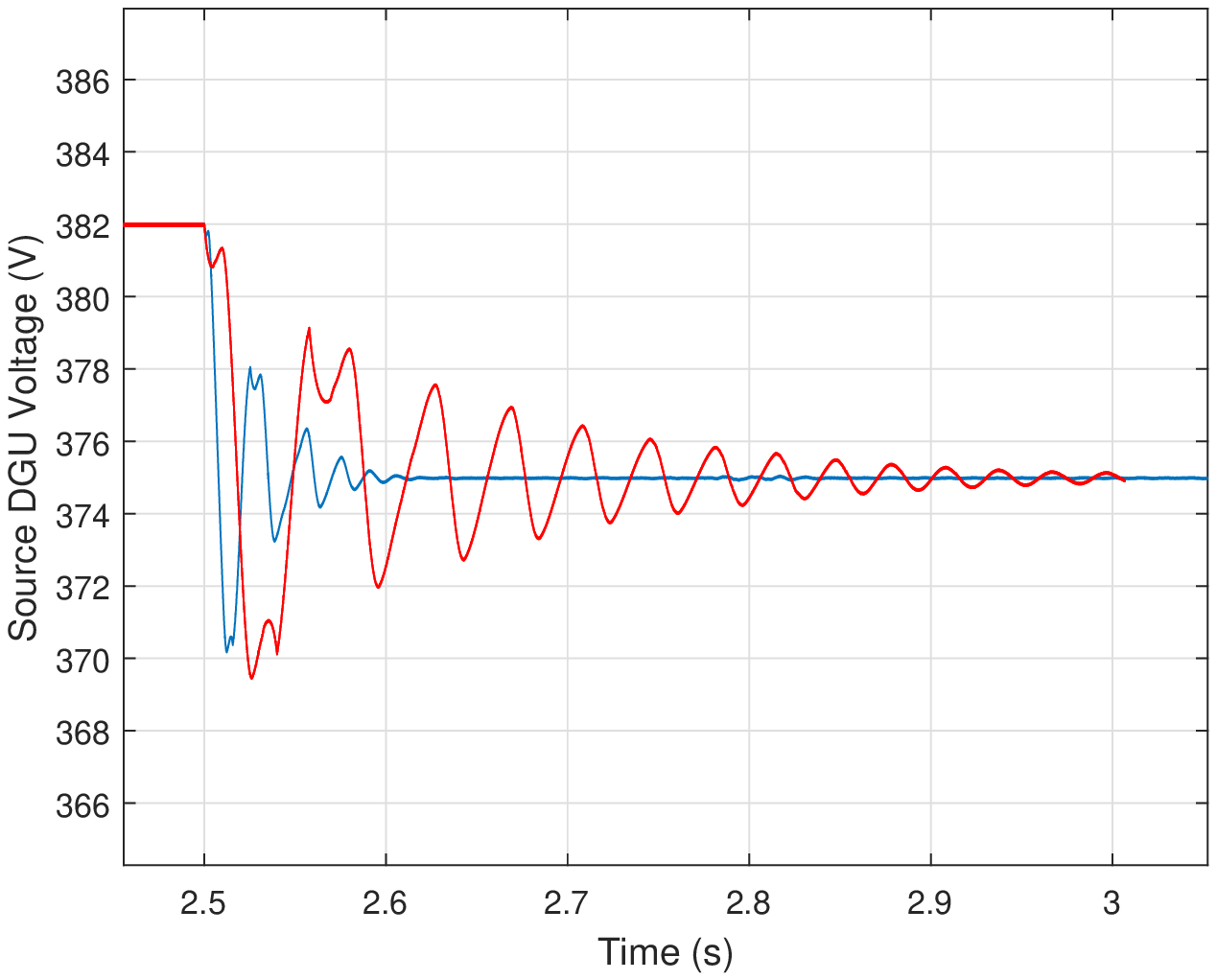}
          \caption{Effect of CPL instability on mG bus voltage: \newline(red) tightly regulated load-side converter; (blue) \newline closed-loop bandwidth detuned.}
   \label{fig:Bode10k}
                 \end{subfigure}
                 \caption{Constant-power load instability.}
                 \label{fig:CPLunstable}
               \end{figure}

  A lightly damped boost converter, stepping 100 V to 382 V is interfaced with a tightly regulated buck converter which powers a resistive load at 48 V. Fig. \ref{fig:CPLunstable}(a) highlights the family of unstable and stable closed-loop eigenvalues as the effective incremental impedance of the boost converter's load is varied from -10 $\Omega$ to 10 $\Omega$.
 Fig.\ref{fig:CPLunstable}(b) shows a poorly damped bus voltage in response to a step from 382 V - 375 V. Fig.\ref{fig:CPLunstable}(b) also shows that by detuning the load-side converter controller, the bus voltage response improves. However, as previously mentioned this would be at the cost of decreased load performance.
   
   \subsection{Stable voltage restoration and equal-current sharing}
   Next, the 6 DGUs of a bus-connected mG, similar to that in Fig.\ref{fig:ConvCct}, are equipped with $\mathcal{C}^{\mathcal{L}_1}_{[i]}$, $i = \{1,...,6\}$, where $\bar{\Sigma}_{[1]}^{\textrm{DGU}}, \bar{\Sigma}_{[2]}^{\textrm{DGU}}, \bar{\Sigma}_{[3]}^{\textrm{DGU}},\bar{\Sigma}_{[4]}^{\textrm{DGU}}$, and $\bar{\Sigma}_{[5]}^{\textrm{DGU}}$ are boost converters while  $\bar{\Sigma}_{[6]}^{\textrm{DGU}}$ is a buck converter. Each DGU provides power to a 380 V DC bus which is loaded with a 5 kW linear resistive load and a 3.8 kW non-linear closed-loop controlled DC motor. Each load is interfaced to the bus via buck converters, where the bus voltage is stepped down to 48 V. System parameters are detailed in Table I of \cite{OKeeffe2018e}.
   
   As shown in section \ref{DesignConsiderations}, $C(s)$ must at least match the smallest relative degree of the plant. Therefore a second-order Butterworth LPF is designed according to (\ref{eqn:L1normCond}), where a frequency sweep from $100-100   \textrm{krads}^{-1}$ is performed.
  \begin{figure}[!htb]    
          \graphicspath{{Images/CPL_Results/}}
          \centering
          \includegraphics[width=8.2cm]{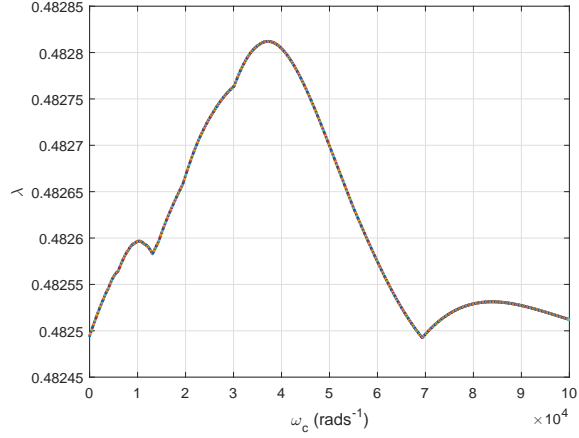}
          \caption{Calculation of appropriate LPF bandwidth $\omega_c$ such that $\lambda < 1$.}
          \label{fig:LPFBand}
          \end{figure} 
  From Fig.\ref{fig:LPFBand}, the bandwidth can be selected arbitrarily. However, from remark 2, in order to allow $\mathbf{Z_{in}(s)}$ to be approximated as a negative-incremental resistance, an upper-bound on the bandwidth of $C(s)$ should be provided. Fig.\ref{fig:CPLunstable}(a) compares the Bode plot of the overall closed-loop load impedance $\mathbf{Z_{in}(s)}$ to the overall open-loop load impedance $\mathbf{Z_{D}(s)}$, i.e. overall high frequency content of CPL loads. In Fig.\ref{fig:CPLunstable}(a), $\mathbf{Z_{in}(s)}$ asymptotically approximates $\mathbf{Z_{D}(s)}$ at 3 krads$^{-1}$ for small CPL levels i.e. each load  equal to 100 W. In Fig.\ref{fig:CPLunstable}(b), $\mathbf{Z_{in}(s)}$ asymptotically approximates $\mathbf{Z_{D}(s)}$ at 50 krads$^{-1}$ for large CPL levels i.e. each load  equal to 10 kW. 
  
  \begin{figure}[!htb]
                 \centering
                 \begin{subfigure}[!htb]{0.52\textwidth}
                   \centering
  \includegraphics[width=\textwidth]{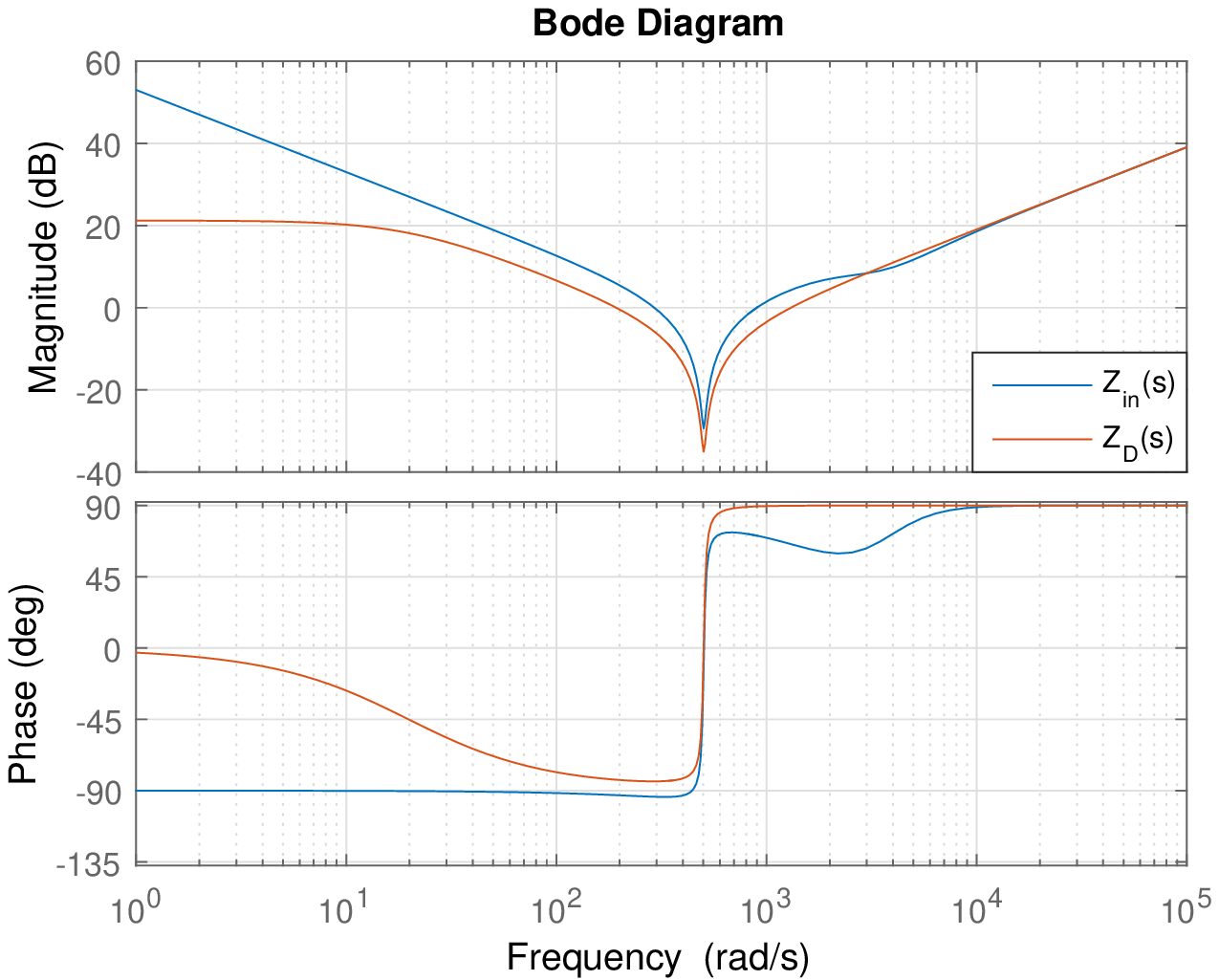}
 \caption{$P_{i}^{\textrm{L}}$ = 100 W.}
                   \label{fig:Bode100}
                 \end{subfigure}\hspace*{\fill}
                 \begin{subfigure}[!htb]{0.52\textwidth}
                   \centering
 \includegraphics[width=\textwidth]{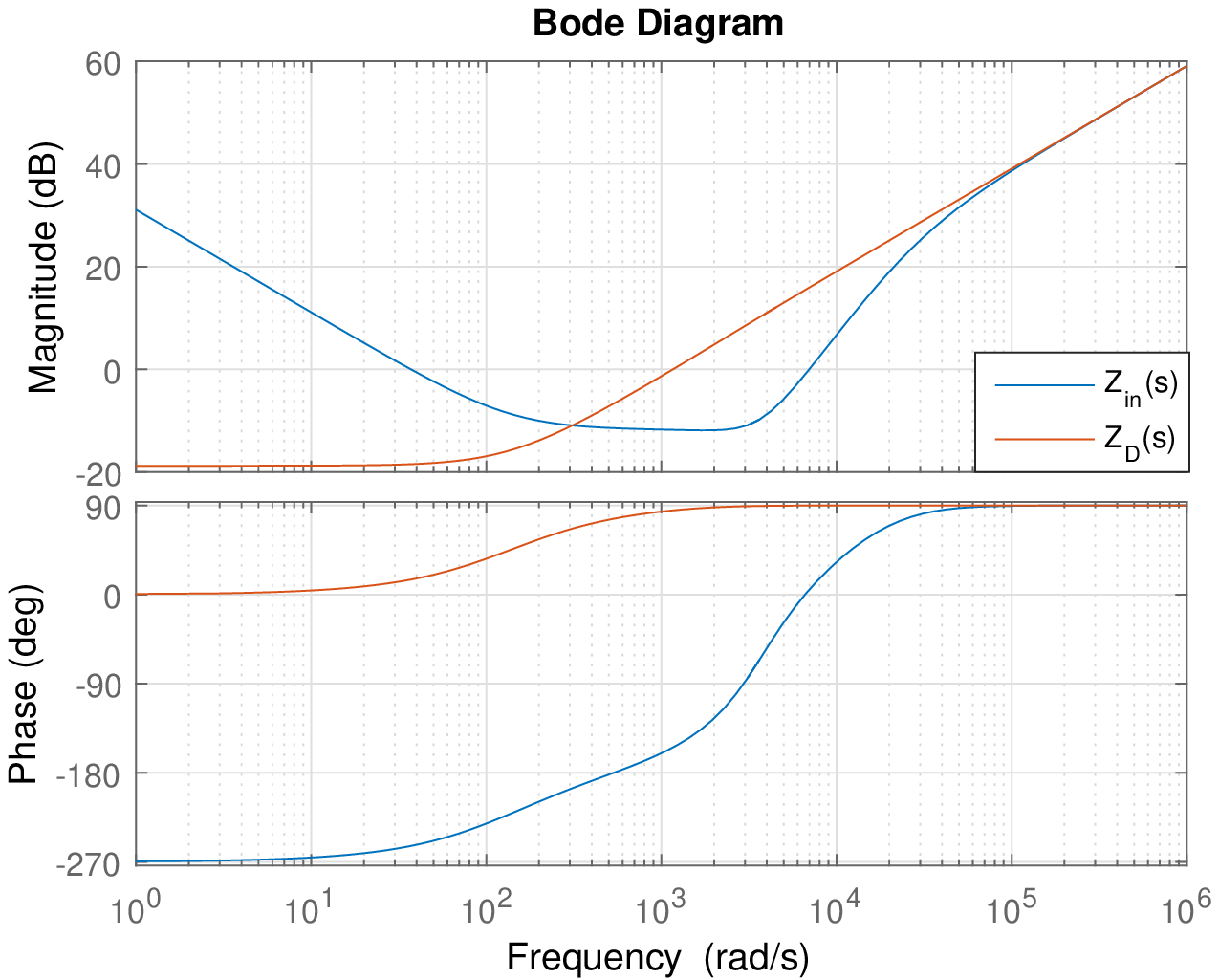}
          \caption{$P_{i}^{\textrm{L}}$ = 10 kW.}
   \label{fig:Bode10k}
                 \end{subfigure}
                 \caption{Bode plots of $\mathbf{Z_{\textrm{in}}(s)}$ under varying $P_{i}^{\textrm{L}}$.}
                 \label{fig:V_Restoration_DGU6PnP}
               \end{figure}

 As a result, the upper-bound is conservatively chosen as 3 krads$^{-1}$. In general, as it is assumed that power levels are unknown, the upper-bound on $C(s)$ should be selected using an \textit{a priori} estimate of the smallest expected load power within the overall mG. Therefore,
 \begin{equation}
 C(s) = \frac{9\times10^6}{s^2+4.243\times10^3s+9\times10^6}
 \end{equation}
  
  The bus-connected mG is powered initially by DGUs $\bar{\Sigma}_{[1]}^{\textrm{DGU}}, \bar{\Sigma}_{[2]}^{\textrm{DGU}}, \bar{\Sigma}_{[3]}^{\textrm{DGU}},\bar{\Sigma}_{[4]}^{\textrm{DGU}}$, and $\bar{\Sigma}_{[5]}^{\textrm{DGU}}$, while $\bar{\Sigma}_{[6]}^{\textrm{DGU}}$ powers a local load on its own. The bus voltage reference $v_{\textrm{bus}}^{\textrm{ref}}$ is set as 380 V. 
  
  \subsubsection{Plug-and-play operation}
  At $t = 8 s$, $\bar{\Sigma}_{[6]}^{\textrm{DGU}}$, which steps an input voltage of 700 V down to 380.5 V, plugs-in. The topology change is represented by the following Laplacian matrices,
  \begin{equation}
  \begin{aligned}
     \mathbb{L}(\mathcal{G})_{t=0s}  \left[\begin{array}{ccccc}
 2  &  -1  &  -1    & 0     &0\\
-1   &  2  &   0    &-1 &    0\\
-1  &   0  &   2    &-1 &    0\\
 0  &  -1  &  -1    & 3 &   -1\\
 0   &  0 &    0    &-1 &    1\\
\end{array} \right] \hspace{2mm} ;\hspace{2mm}  \mathbb{L}(\mathcal{G})_{t=8s}   \left[\begin{array}{cccccc}
 3  &  -1  &  -1    & 0   &0  &-1\\
-1   &  2  &   0    &-1 &  0&  0\\
-1  &   0  &   2    &-1 &   0& 0\\
 0  &  -1  &  -1    & 3 &   0&-1\\
 0   &  0 &    0    &-1 &    2& -1\\
 -1   &  0 &    0    &0 &    -1& 2\\
\end{array} \right]
      \end{aligned}
      \label{eqn:Lap}
  \end{equation}
  
  Fig.\ref{fig:V_Restoration_DGU6PnP}(a) shows the response of all the DGUs when a fully-connected communications network is used. Fig.\ref{fig:V_Restoration_DGU6PnP}(b) shows the response when a sparse connected communications network with optimal link redundancy is used.
  
  \begin{figure}[!htb]
                 \centering
                 \begin{subfigure}[!htb]{0.52\textwidth}
                   \centering
  \includegraphics[width=\textwidth]{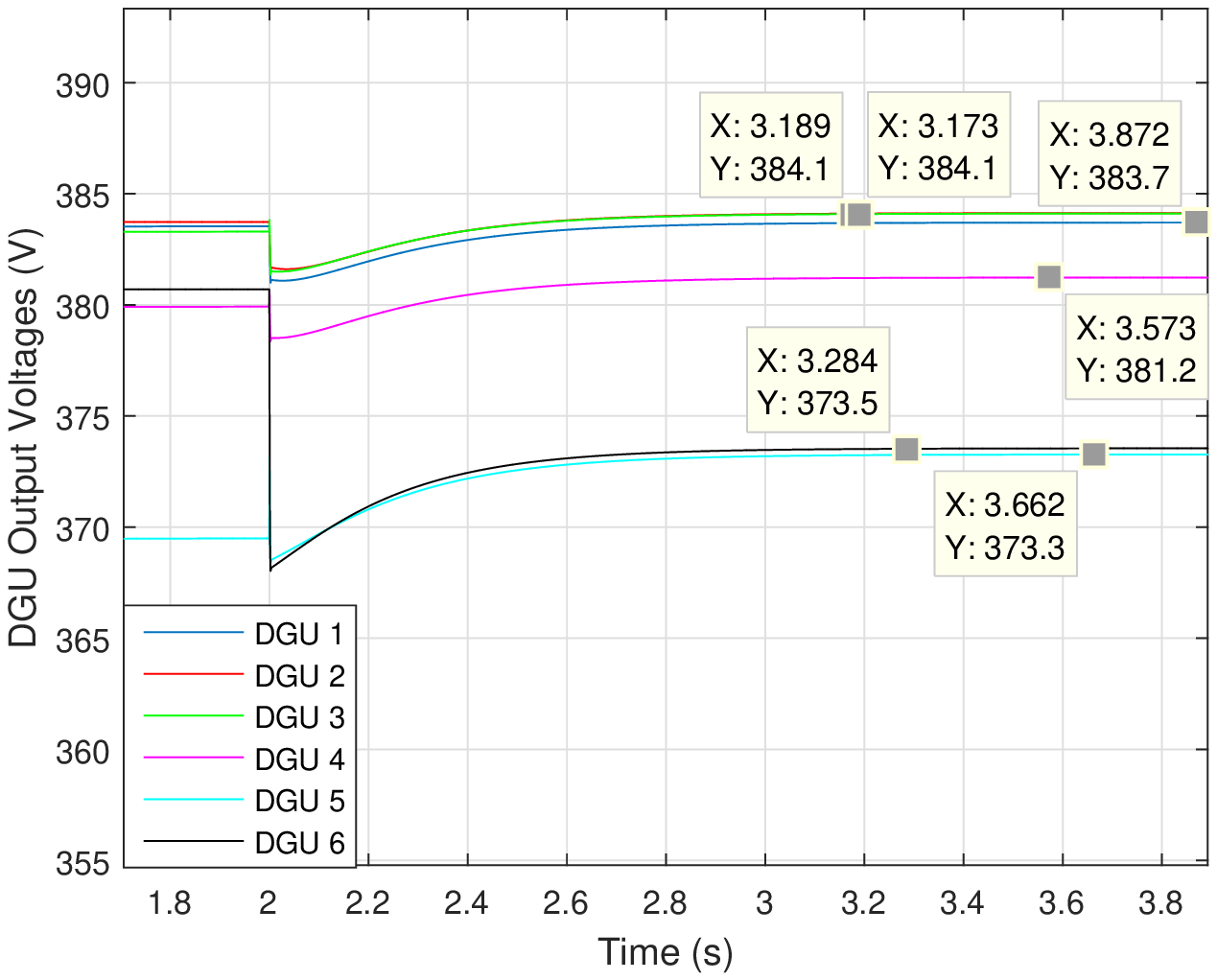}
 \caption{Fully-connected communications network.}
                   \label{fig:DGU1PnPDGU6}
                 \end{subfigure}\hspace*{\fill}
                 \begin{subfigure}[!htb]{0.52\textwidth}
                   \centering
 \includegraphics[width=\textwidth]{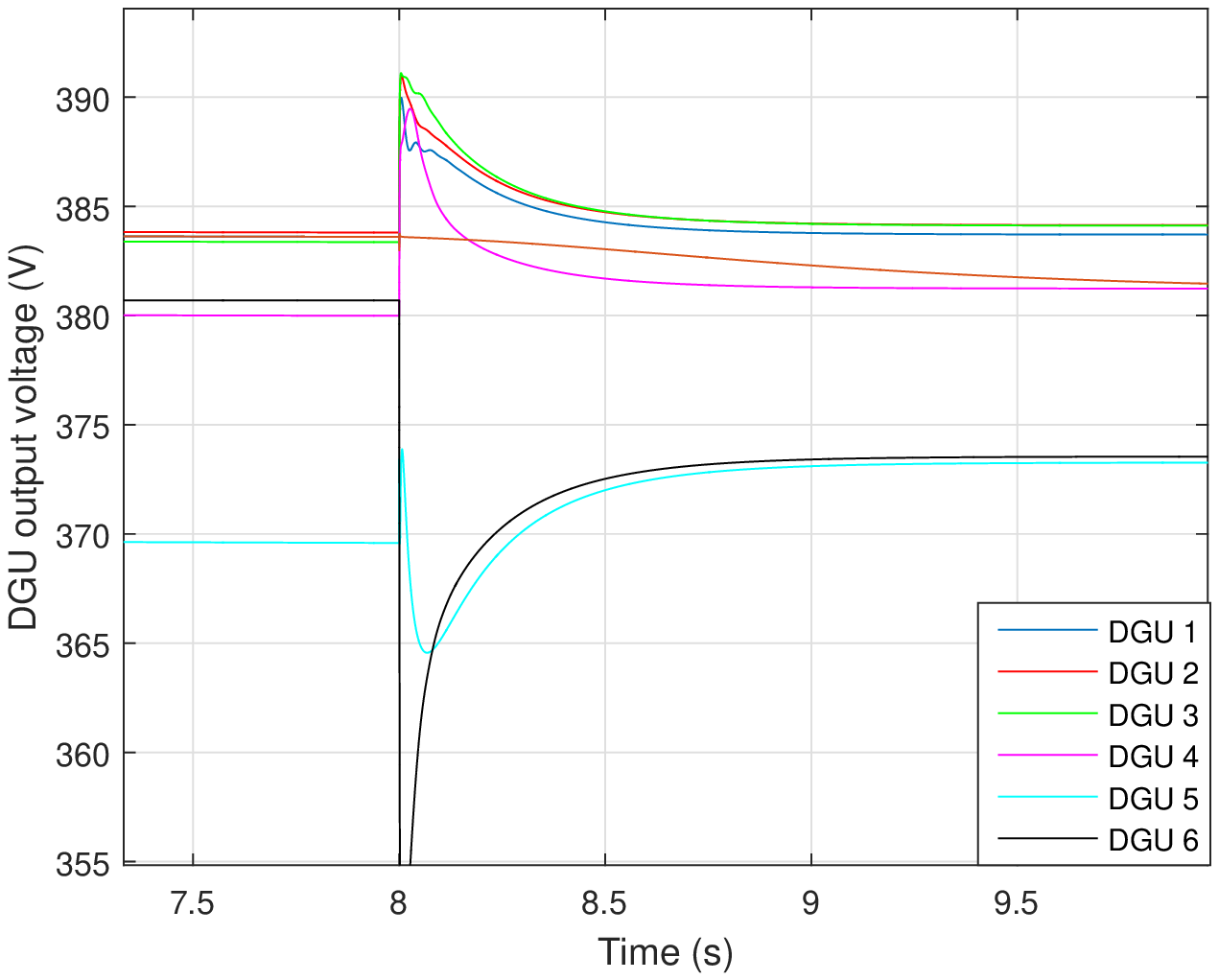}
          \caption{Sparse communications network.}
   \label{fig:DGU5PnPDGU6}
                 \end{subfigure}
                 \caption{Voltage restoration of the microgrid bus voltage when $\bar{\Sigma}_{[6]}^{\textrm{DGU}}$ plugs-in.}
                 \label{fig:V_Restoration_DGU6PnP}
               \end{figure}

          Here, it is clear that all DGUs restore their output voltages such that their average equals 380 V. The secondary control response is adequately fast with 0.7 s settling times and the sparse network compares well with the fully-connected network.
          
          Fig.\ref{fig:Load_Share_DGU6PnP} shows the maintenance of equal-current sharing during the plug-in test. Again, a fully-connected network and the sparse network are compared.
             \begin{figure}[!htb]
                 \centering
                 \begin{subfigure}[!htb]{0.52\textwidth}
                   \centering
  \includegraphics[width=\textwidth]{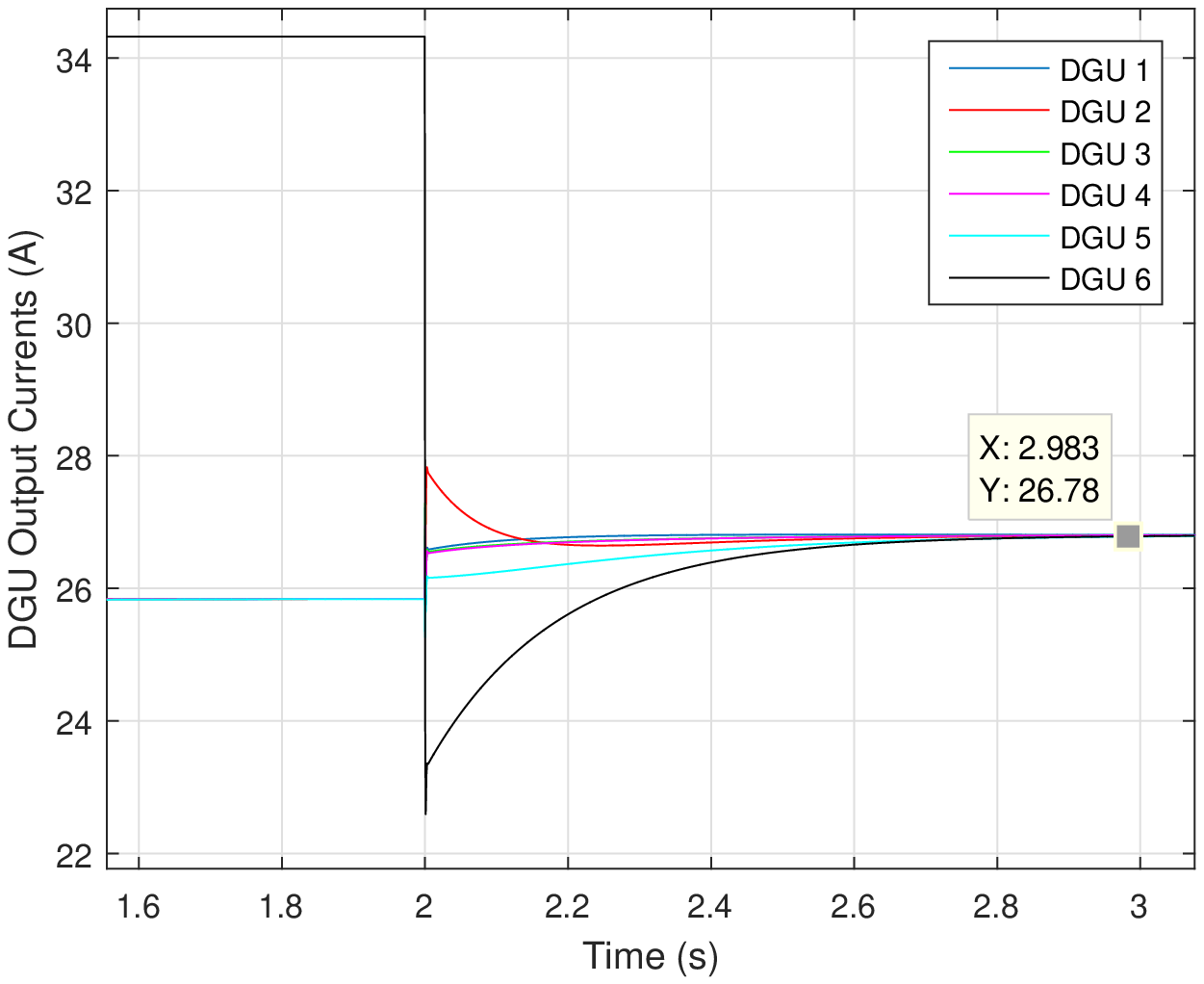}
 \caption{Fully-connected communications network.}
                   \label{fig:DGU1PnPDGU6}
                 \end{subfigure}\hspace*{\fill}
                 \begin{subfigure}[!htb]{0.52\textwidth}
                   \centering
 \includegraphics[width=\textwidth]{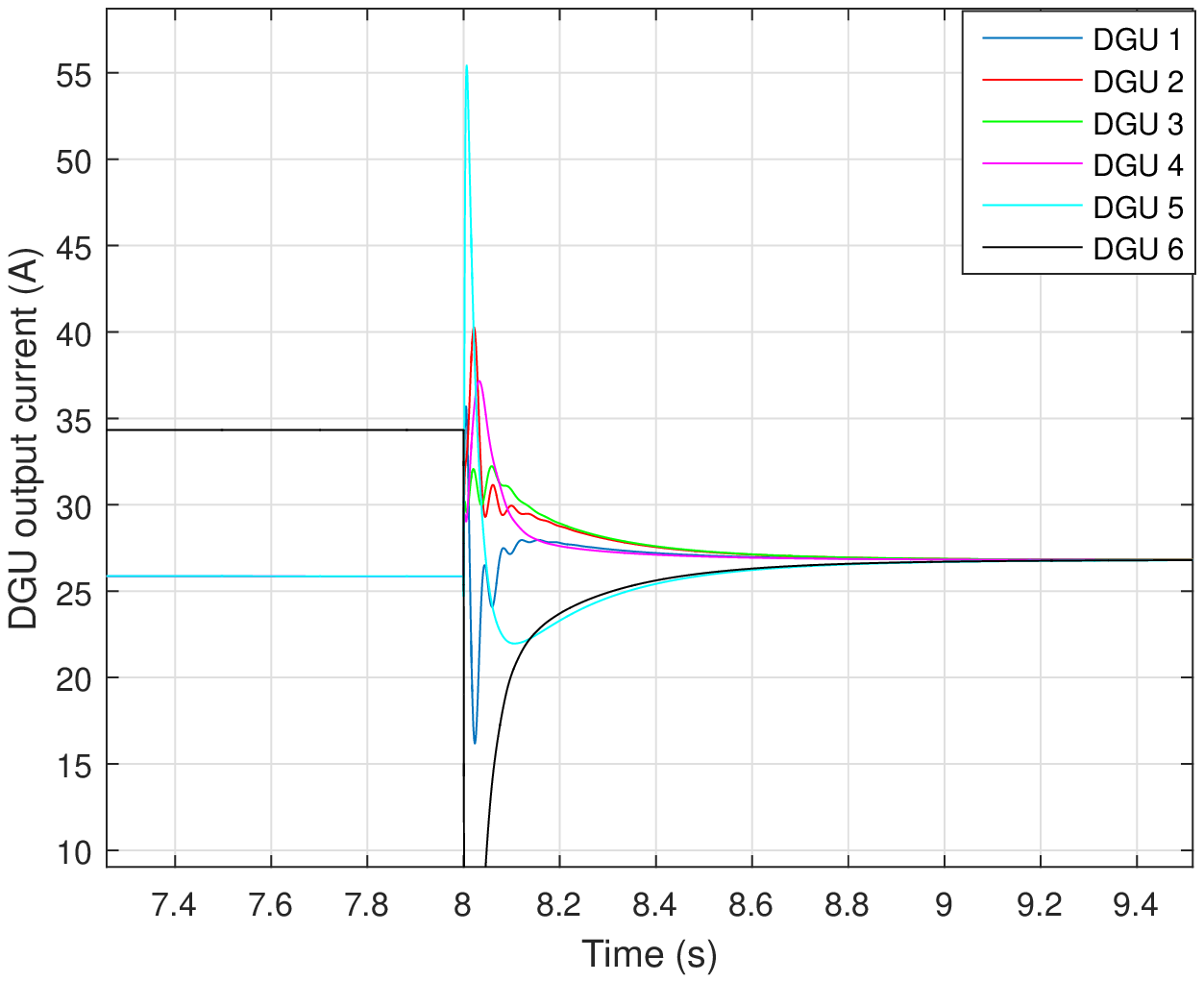}
          \caption{Sparse communications network.}
   \label{fig:DGU5PnPDGU6}
                 \end{subfigure}
                 \caption{Equal-current sharing maintained when $\bar{\Sigma}_{[6]}^{\textrm{DGU}}$ plugs-in.}
                 \label{fig:Load_Share_DGU6PnP}
                    \end{figure} 
                    
                    When $\bar{\Sigma}_{[6]}^{\textrm{DGU}}$ joins the network, all DGU output current injections quickly converge to a current of 26.78 A. Though steady-state performance is similar for both networks with a settling time of approximately 0.6 s, the transient response of $\bar{\Sigma}_{[5]}^{\textrm{DGU}}$ in particular has a large 55$\%$ overshoot. As a result $\bar{\Sigma}_{[6]}^{\textrm{DGU}}$ undershoots by the same amount in order to satisfy (\ref{eqn:shareI}). 
                    
                    The convergence rates at start-up, i.e. $t= 0 s$, and at $t = 8 s$ when the topology reconfigures i.e. $\mathbb{L}(\mathcal{G})$ changes, are plotted below.
                    
                    \begin{figure}[!htb]
                 \centering
                 \begin{subfigure}[!htb]{0.52\textwidth}
                   \centering
  \includegraphics[width=\textwidth]{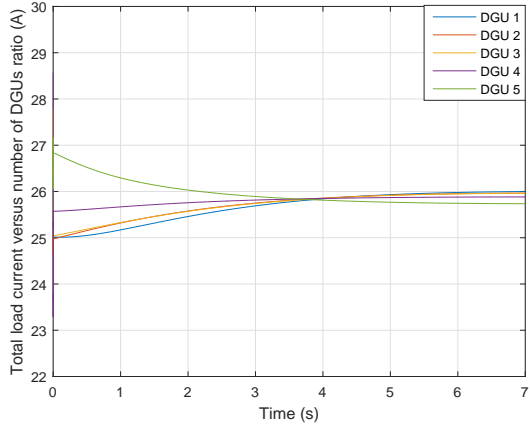}
 \caption{Convergence of DGU current estimates at start-up.}
                   \label{fig:DGU1PnPDGU6}
                 \end{subfigure}\hspace*{\fill}
                 \begin{subfigure}[!htb]{0.52\textwidth}
                   \centering
 \includegraphics[width=\textwidth]{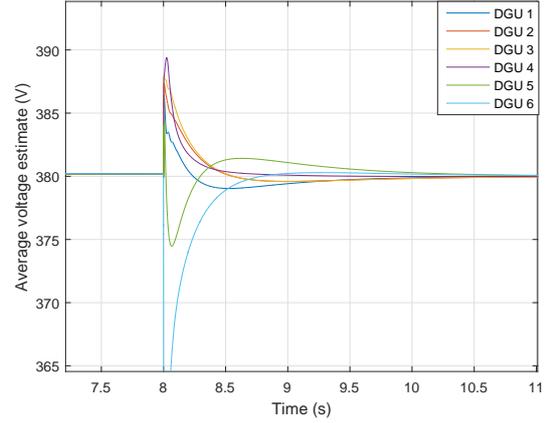}
          \caption{Sparse communications network.}
   \label{fig:DGU5PnPDGU6}
                 \end{subfigure}
                 \caption{Convergence of DGU current and voltage estimates.}
                 \label{fig:Load_Share_DGU6PnP}
                    \end{figure} 
                 
                 \subsubsection{Link-failure resiliency}
                 Finally, the system's resiliency to communication link failures is investigated. At $t = 14 s$, though the topology of the physical system remains the same when $\bar{\Sigma}_{[6]}^{\textrm{DGU}}$ plugs-in at $t = 8 s$, the topology of the communication network changes. Edges $\varepsilon_{12}$ and $\varepsilon_{13}$ fail.  $\bar{\Sigma}_{[1]}^{\textrm{DGU}}$ can only exchange information with $\bar{\Sigma}_{[6]}^{\textrm{DGU}}$. The Laplace changes from $\mathbb{L}(\mathcal{G})_{t=8s}$ in (\ref{eqn:Lap}) to,
                 \begin{equation}
                 \begin{aligned}
      \mathbb{L}(\mathcal{G})_{t=14s}   \left[\begin{array}{cccccc}
 1  &  0  &  0    & 0   &0  &-1\\
0&  0  &   1    &-1 &  0&  0\\
 0 &   0  &   1    &-1 &   0& 0\\
 0  &  -1  &  -1    & 2 &   -1&0\\
 0   &  0 &    0    &-1 &    2& -1\\
 -1   &  0 &    0    &0 &    -1& 2\\
\end{array} \right]
      \end{aligned}
      \label{eqn:LapFail}
  \end{equation}
  The response to the change in communication topology is shown below.
  \begin{figure}[!htb]
                 \centering
                 \begin{subfigure}[!htb]{0.52\textwidth}
                   \centering
  \includegraphics[width=\textwidth]{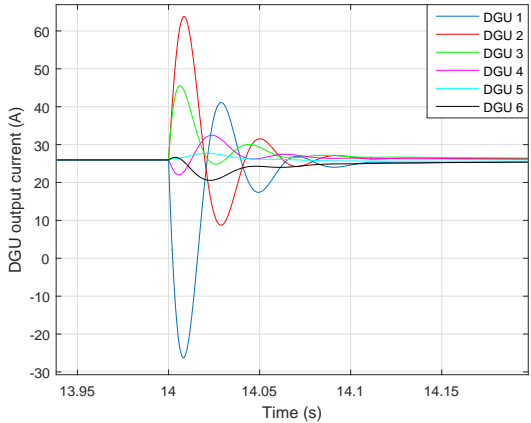}
 \caption{DGU output current response.}
                   \label{fig:DGU1PnPDGU6}
                 \end{subfigure}\hspace*{\fill}
                 \begin{subfigure}[!htb]{0.52\textwidth}
                   \centering
 \includegraphics[width=\textwidth]{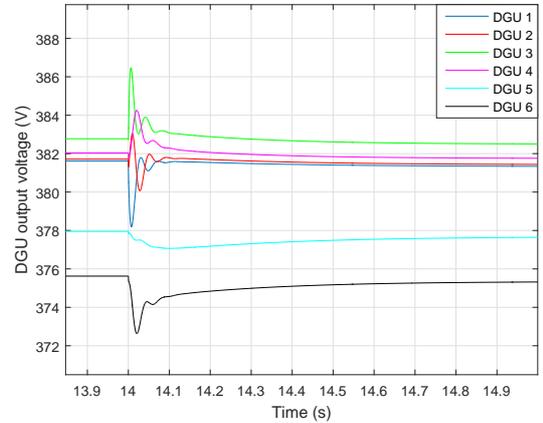}
          \caption{DGU output voltage response.}
   \label{fig:DGU5PnPDGU6}
                 \end{subfigure}
                 \caption{Output current and voltage responses when communications links $\varepsilon_{12}$ and $\varepsilon_{13}$ fail.}
                 \label{fig:LinkRez}
                    \end{figure} 
Fig.\ref{fig:LinkRez} shows the system's fault tolerant capability. As long as each node in the communication network has a neighbour such that information can be exchanged throughout the system either directly or indirectly then system irrespective of communication faults. Ultimately, the topology change, reflected by the Laplacian matrices, only affects the transient response. In general, the less nodes connected the slower the response and the larger over/under shoots can be, as seen in Fig.\ref{fig:LinkRez}(a).
                    
  \section{Conclusion}
  This paper develops a novel scalable and purely adaptive $\mathcal{L}_1$AC architecture to stabilise DGUs in a bus-connected DC mG with CPLs. The architecture draws on a control orientated load-connected model which enables scalability and treats unknown loads and exogenous inputs. Due to the adaptive nature of the architecture, a bus-connected topology does not need to be mapped into this load-connected topology, as long as the bus-connected parameters exist within the uncertainty bound.
  
  The destabilising current disturbances induced by the CPL at each DGU terminal is approximated by its low-frequency content due to the LPF of the  $\mathcal{L}_1$AC architecture. Necessary and sufficient stability conditions are provided for DC-DC boost and buck converters using a nominally expected load power. Subsequently, the smallest \textit{a priori} expected load total is used to determine an upper-bound for designing the bandwidth of the architecture's LPF. As a result, very fast asymptotic convergence of local state and estimation errors is theoretically proven and demonstrated with a bus-connected mG. For overall system stability of the primary level, a conservative, offline approach is used.
  
   The proposed adaptive primary controllers are cascaded with distributed consensus-based secondary controllers where global knowledge of the average bus voltage and load current are estimated using limited local information. Asymptotic stability of the hierarchical control system is shown using Lyapunov functions. Voltage restoration and equal-current sharing responses are shown to be fast and stable during PnP and communication link failure tests.  
   
\bibliography{arXiv_UKACC_Paper}

\begin{thebibliography}{10}
\providecommand{\url}[1]{#1}
\csname url@samestyle\endcsname
\providecommand{\newblock}{\relax}
\providecommand{\bibinfo}[2]{#2}
\providecommand{\BIBentrySTDinterwordspacing}{\spaceskip=0pt\relax}
\providecommand{\BIBentryALTinterwordstretchfactor}{4}
\providecommand{\BIBentryALTinterwordspacing}{\spaceskip=\fontdimen2\font plus
\BIBentryALTinterwordstretchfactor\fontdimen3\font minus
  \fontdimen4\font\relax}
\providecommand{\BIBforeignlanguage}[2]{{%
\expandafter\ifx\csname l@#1\endcsname\relax
\typeout{** WARNING: IEEEtran.bst: No hyphenation pattern has been}%
\typeout{** loaded for the language `#1'. Using the pattern for}%
\typeout{** the default language instead.}%
\else
\language=\csname l@#1\endcsname
\fi
#2}}
\providecommand{\BIBdecl}{\relax}
\BIBdecl

\bibitem{Anuradha2013}
M.~A. Anuradha and A.~Massoud, \emph{{IEEE Vision for Smart Grid Controls: 2030
  and Beyond: Roadmap}}.\hskip 1em plus 0.5em minus 0.4em\relax IEEE CSS, 2013.

\bibitem{Andreasson2016}
M.~Andreasson, D.~V. Dimarogonas, H.~Sandberg, and K.~H. Johansson,
  ``{Distributed controllers for multiterminal HVDC transmission systems},''
  \emph{IEEE Transactions on Control of Network Systems}, vol.~4, no.~3, pp.
  564--574, 2017.

\bibitem{Stoustrup2009}
J.~{\~{A}}. Stoustrup, ``{Plug {\&} Play Control : Control Technology Towards
  New Challenges},'' \emph{European Journal of Control}, vol.~15, no. 3-4, pp.
  311--330, 2009.

\bibitem{Shafiee2014}
Q.~Shafiee, T.~Dragicevic, J.~C. Vasquez, and J.~M. Guerrero, ``{Modeling,
  stability analysis and active stabilization of multiple DC-microgrid
  clusters},'' in \emph{ENERGYCON 2014 - IEEE International Energy Conference},
  2014, pp. 1284--1290.

\bibitem{Riccobono2014}
A.~Riccobono\vspace{0mm} and E.~Santi, ``Comprehensive review of stability
  criteria for dc power distribution systems,'' \emph{IEEE Transactions on
  Industry Applications}, vol.~50, no.~5, pp. 3525--3535, 2014.

\bibitem{Vasquez2016}
T.~Dragicevic, X.~Lu, J.~C. Vasquez, and J.~M. Guerrero, ``{DC Microgrids -
  Part I: A Review of Control Strategies and Stabilization Techniques},''
  \emph{IEEE Transactions on Power Electronics}, vol.~31, no.~7, pp.
  4876--4891, 2016.

\bibitem{Riccobono2012}
A.~Riccobono and E.~Santi, ``A novel passivity-based stability criterion (pbsc)
  for switching converter dc distribution systems,'' in \emph{Applied Power
  Electronics Conference and Exposition (APEC), 2012 Twenty-Seventh Annual
  IEEE}.\hskip 1em plus 0.5em minus 0.4em\relax IEEE, 2012, pp. 2560--2567.

\bibitem{Riccobono2013a}
A.~Riccobono\vspace{0mm} and E.~Santi, ``Stability analysis of an all-electric
  ship mvdc power distribution system using a novel passivity-based stability
  criterion,'' in \emph{Electric Ship Technologies Symposium (ESTS), 2013
  IEEE}.\hskip 1em plus 0.5em minus 0.4em\relax IEEE, 2013, pp. 411--419.

\bibitem{Riverso2015}
S.~Riverso, F.~Sarzo, and G.~Ferrari-Trecate, ``{Plug-and-Play Voltage and
  Frequency Control of Islanded Microgrids with Meshed Topology},'' \emph{IEEE
  Transactions on Smart Grid}, vol.~6, no.~3, pp. 1176--1184, 2015.

\bibitem{Riverso2017a}
S.~Riverso, M.~Tucci, J.~C. Vasquez, J.~M. Guerrero, and G.~Ferrari-Trecate,
  ``{Stabilizing plug-and-play regulators and secondary coordinated control for
  AC islanded microgrids with bus-connected topology},'' \emph{Applied Energy},
  no. August, pp. 1--21, 2017.

\bibitem{Tucci2016c}
M.~Tucci, S.~Riverso, J.~C. Vasquez, J.~M. Guerrero, and G.~Ferrari-Trecate,
  ``{A Decentralized Scalable Approach to Voltage Control of DC Islanded
  Microgrids},'' \emph{IEEE Transactions on Control Systems Technology},
  vol.~24, no.~6, pp. 1965--1979, 2016.

\bibitem{Tucci2017g}
M.~Tucci, L.~Meng, J.~M. Guerrero, and G.~Ferrari-Trecate, ``{Plug-and-play
  control and consensus algorithms for current sharing in DC microgrids},'' in
  \emph{IFAC-PapersOnLine}, vol.~50, 2016, pp. 1--23.


\bibitem{Tucci2016}
\BIBentryALTinterwordspacing
M.~Tucci, L.~Meng, J.~M. Guerrero, and G.~Ferrari-Trecate, ``{Consensus
  algorithms and plug-and-play control for current sharing in DC microgrids},''
  \emph{arXiv}, pp. 1--23, 2016. [Online]. Available:
  \url{http://arxiv.org/abs/1603.03624}
\BIBentrySTDinterwordspacing


\bibitem{Sadabadi2017a}
M.~S. Sadabadi, Q.~Shafiee, and A.~Karimi, ``{Plug-and-Play Robust Voltage
  Control of DC Microgrids},'' \emph{IEEE Transactions on Smart Grid}, pp.
  1--1, 2017.

\bibitem{Han2017}
\BIBentryALTinterwordspacing
R.~Han, M.~Tucci, R.~Soloperto, A.~Martinelli, G.~Ferrari-Trecate, and J.~M.
  Guerrero, ``{Hierarchical Plug-and-Play Voltage/Current Controller of DC
  microgrid with Grid-Forming/Feeding modules: Line-independent Primary
  Stabilization and Leader-based Distributed Secondary Regulation},'' no. July,
  2017. [Online]. Available: \url{http://arxiv.org/abs/1707.07259}
\BIBentrySTDinterwordspacing

\bibitem{D??rfler2013}
F.~Dorfler and F.~Bullo, ``{Kron reduction of graphs with applications to
  electrical networks},'' \emph{IEEE Transactions on Circuits and Systems I:
  Regular Papers}, vol.~60, no.~1, pp. 150--163, 2013.

\bibitem{Tucci2017}
M.~Tucci, A.~Floriduz, S.~Riverso, and G.~Ferrari-Trecate, ``Plug-and-play
  control of ac islanded microgrids with general topology,'' in \emph{Control
  Conference (ECC), 2016 European}.\hskip 1em plus 0.5em minus 0.4em\relax
  IEEE, 2016, pp. 1493--1500.

\bibitem{Katiraei2017}
F.~Katiraei, A.~Zamani, and R.~Masiello, ``{Microgrid Control Systems},''
  \emph{IEEE Power and Energy Magazine}, vol.~15, no.~4, pp. 116--112, 2017.

\bibitem{Nasirian2014a}
V.~Nasirian, A.~Davoudi, F.~L. Lewis, and J.~M. Guerrero, ``{Distributed
  adaptive droop control for DC distribution systems},'' \emph{IEEE
  Transactions on Energy Conversion}, vol.~29, no.~4, pp. 944--956, 2014.

\bibitem{Shafiee2014b}
Q.~Shafiee, T.~Dragicevic, F.~Andrade, J.~C. Vasquez, and J.~M. Guerrero,
  ``Distributed consensus-based control of multiple dc-microgrids clusters,''
  in \emph{Industrial Electronics Society, IECON 2014-40th Annual Conference of
  the IEEE}.\hskip 1em plus 0.5em minus 0.4em\relax IEEE, 2014, pp. 2056--2062.

\bibitem{Josep2014}
X.~Lu, K.~Sun, J.~M. Guerrero, J.~C. Vasquez, and L.~Huang, ``{State-of-Charge
  Balance Using Adaptive Droop Control for Distributed Energy Storage Systems
  in DC Microgrid Applications},'' \emph{IEEE Trans. Ind. Electron.}, vol.~61,
  pp. 2804--2815, 2014.

\bibitem{Vu2017}
T.~V. Vu, D.~Perkins, F.~Diaz, D.~Gonsoulin, C.~S. Edrington, and
  T.~El-Mezyani, ``{Robust adaptive droop control for DC microgrids},''
  \emph{Electric Power Systems Research}, vol. 146, pp. 95--106, 2017.

\bibitem{Lu2014}
X.~Lu, J.~M. Guerrero, K.~Sun, and J.~C. Vasquez, ``{An improved droop control
  method for dc microgrids based on low bandwidth communication with dc bus
  voltage restoration and enhanced current sharing accuracy},'' \emph{IEEE
  Transactions on Power Electronics}, vol.~29, no.~4, pp. 1800--1812, 2014.

\bibitem{Simpson2013}
J.~W. Simpson-Porco, F.~D{\"o}rfler, and F.~Bullo, ``Synchronization and power
  sharing for droop-controlled inverters in islanded microgrids,''
  \emph{Automatica}, vol.~49, no.~9, pp. 2603--2611, 2013.

\bibitem{Zhao2015}
J.~Zhao and F.~D{\"o}rfler, ``Distributed control and optimization in dc
  microgrids,'' \emph{Automatica}, vol.~61, pp. 18--26, 2015.

\bibitem{L12010}
C.~Cao and Hovakimyan, \emph{{L1 Adaptive Control Theory: Guaranteed Robustness
  with Fast Adaptation}}.\hskip 1em plus 0.5em minus 0.4em\relax Society for
  Industrial and Applied Mathematics, 2010.

\bibitem{Gibson2012}
T.~E. Gibson, A.~M. Annaswamy, and E.~Lavretsky, ``{Improved Transient Response
  in Adaptive Control Using Projection Algorithms and Closed Loop Reference
  Models},'' \emph{AIAA Guidance, Navigation, and Control Conference}, no.
  August, pp. 1--13, 2012.

\bibitem{OKeeffe2018e}
\BIBentryALTinterwordspacing
D.~O'Keeffe\vspace{0mm}, S.~Riverso, L.~Albiol-Tendillo, and G.~Lightbody,
  ``{Voltage Control of DC Islanded Microgrids: Scalable Decentralised L1
  Adaptive Controllers},'' 2018. [Online]. Available: \url{arXiv preprint
  arXiv:1801.04508}
\BIBentrySTDinterwordspacing

\bibitem{OKeeffe2018c}
\BIBentryALTinterwordspacing
D.~O'Keeffe, S.~Riverso\vspace{0mm}, L.~Albiol-Tendillo, and G.~Lightbody, ``{A
  Distributed Scalable Architecture using L1 Adaptive Controllers for Primary
  Voltage Control of DC Microgrids},'' 2018. [Online]. Available: \url{arXiv
  preprint arXiv:1801.06484}
\BIBentrySTDinterwordspacing

\bibitem{Lu2014a}
X.~Lu, K.~Sun, L.~Huang, J.~M. Guerrero, J.~C. Vasquez, and Y.~Xing, ``{Virtual
  impedance based stability improvement for DC microgrids with constant power
  loads},'' \emph{2014 IEEE Energy Conversion Congress and Exposition, ECCE
  2014}, vol.~6, no.~6, pp. 2670--2675, 2014.

\bibitem{Arcidiacono2012}
V.~Arcidiacono, A.~Monti, and G.~Sulligoi, ``Generation control system for
  improving design and stability of medium-voltage dc power systems on ships,''
  \emph{IET Electrical Systems in Transportation}, vol.~2, no.~3, pp. 158--167,
  2012.

\bibitem{Luis2017}
L.~Herrera and J.~Wang, ``Stability analysis and controller design of dc
  microgrids with constant power loads,'' in \emph{Applied Power Electronics
  Conference and Exposition (APEC), 2015 IEEE}.\hskip 1em plus 0.5em minus
  0.4em\relax IEEE, 2015, pp. 691--696.

\bibitem{OKeeffe2017a}
D.~O'Keeffe, S.~Riverso, L.~Albiol-Tendillo, and G.~Lightbodyt\vspace{0mm},
  ``{Distributed Hierarchical Droop Control of Boost Converters in DC
  Microgrids},'' \emph{28th IEEE Irish Signals and Systems Conference}, pp.
  1--6, 2017.

\bibitem{Gregory2010}
I.~Gregory, E.~Xargay, C.~Cao, and N.~Hovakimyan, ``{Flight Test of an L1
  Adaptive Controller on the NASA AirSTAR Flight Test Vehicle},'' \emph{AIAA
  Guidance, Navigation, and Control Conference}, pp. 1--31, 2010.

\bibitem{Ackerman2017}
K.~A. Ackerman, E.~Xargay, R.~Choe, N.~Hovakimyan, C.~M. Cotting, R.~B.
  Jeffrey, M.~P. Blackstun, P.~T. Fulkerson, T.~R. Lau, and S.~S. Stephens,
  ``{Evaluation of an L1 Adaptive Flight Control Law on Calspans
  Variable-Stability Learjet},'' \emph{Journal of Guidance, Control, and
  Dynamics}, pp. 1--10, 2017.

\bibitem{Svendsen2012}
C.~H. Svendsen, N.~O. Holck, R.~Galeazzi, and M.~Blanke, ``{L1 adaptive
  manoeuvring control of unmanned high-speed water craft},'' \emph{IFAC
  Proceedings Volumes (IFAC-PapersOnline)}, vol.~9, no. PART 1, pp. 144--151,
  2012.

\bibitem{Michini2009}
B.~Michini and J.~P. How, ``{L1 adaptive control for indoor autonomous
  vehicles: Design process and flight testing},'' \emph{Proceeding of AIAA
  Guidance, Navigation, and Control Conference}, no. August, pp. 1--15, 2009.

\bibitem{Yoo2010}
S.~Yoo, C.~Cao, and N.~Hovakimyan, ``{Decentralised L1 adaptive control for
  large-scale non-linear systems with interconnected unmodelled dynamics},''
  \emph{IET Control Theory {\&} Applications}, vol.~4, no.~10, pp. 1972--1988,
  2010.

\bibitem{DeL1AC}
S.~Yoo, N.~Hovakimyan, and C.~Cao, \emph{Decentralized L1 adaptive control for
  large-scale systems with unknown time-varying interaction parameters}, 10
  2010, pp. 5590--5595.

\bibitem{Kumaresan2016}
G.~Kumaresan and A.~Kale, ``{Application of L1 adaptive controller for the
  design of a novel decentralized leader follower formation algorithm},''
  \emph{IFAC-PapersOnLine}, vol.~49, no.~1, pp. 706--711, 2016.

\bibitem{Lavretsky2011}
\BIBentryALTinterwordspacing
E.~Lavretsky and T.~E. Gibson, ``{Projection Operator in Adaptive Systems},''
  \emph{arXiv preprint, arXiv:1112.4232}, 2011. [Online]. Available:
  \url{http://arxiv.org/abs/1112.4232}
\BIBentrySTDinterwordspacing

\bibitem{Kurucs2015}
B.~Kurucs, ``{State Space Control of Quadratic Boost Converter using LQR and
  LQG approaches},'' \emph{2015 Intl Conference on Optimization of Electrical
  {\&} Electronic Equipment}, no.~2, pp. 642--648, 2015.

\end{thebibliography}
\end{document}